%% file: mink2017_gmcmf_arxiv.tex
\documentclass[useAMS,twocolumn]{aastex6}

\usepackage{amsmath,amssymb}
\usepackage{natbib}
\usepackage{epsfig}
\usepackage{times}
\usepackage{rotate}
\usepackage{hyperref}
\usepackage{color,url}
\usepackage{subfigure}
\usepackage{cprotect}
\usepackage{xspace}
\usepackage{enumerate}
\usepackage{fancyhdr}
\usepackage{appendix}
\usepackage[pass]{geometry}

\newcommand{\simgt}{\lower.5ex\hbox{$\; \buildrel > \over \sim \;$}}
\newcommand{\simlt}{\lower.5ex\hbox{$\; \buildrel < \over \sim \;$}}

\newcommand{\myemail}{masato.kobayashi@nagoya-u.jp}

\newcommand{\eg}{{\it e.g.}\xspace}
\newcommand{\cf}{{\it c.f.}\xspace}
\newcommand{\ie}{{\it i.e.}\xspace}
\newcommand{\etc}{{\it etc.}}

\newcommand{\msun}{\mbox{${\rm M_{\odot}}$}\xspace}

\newcommand{\mytf}{T_{\rm f}}
\newcommand{\mytsg}{T_{\rm sg}}
\newcommand{\mytd}{T_{\rm d}}
\newcommand{\mytsb}{T_{*}}
\newcommand{\mytsd}{T_{\rm dest}}
\newcommand{\mytffid}{T_{\rm f,fid}}

\newcommand{\mytcol}{T_{\rm col}}

\newcommand{\mytsteady}{T_{\rm steady}}
\newcommand{\myratio}{\mytf/\mytd}
\newcommand{\mtrunc}{m_{\rm trunc}}
\newcommand{\myrec}{\varepsilon_{\rm res}}
\newcommand{\myrecsteady}{\varepsilon_{\rm res,std}}

\newcommand{\mcrit}{m_{\rm crit}}

\newcommand{\myfin}{F_{\rm in}}
\newcommand{\myfout}{F_{\rm out}}
\newcommand{\mydotmg}{\dot{M}_{\rm grow}}
\newcommand{\mydotmd}{\dot{M}_{\rm disp}}
\newcommand{\mymmin}{m_{\rm min}}
\newcommand{\mymmax}{m_{\rm max}}

\begin{document}


    \title{Evolutionary Description of Giant Molecular Cloud Mass Functions on Galactic Disks}
    \shorttitle{Time Evolution of GMC Mass Functions}




    \author{Masato I.N. Kobayashi\altaffilmark{1},
            Shu-ichiro Inutsuka\altaffilmark{1},
            Hiroshi Kobayashi\altaffilmark{1},
            and
            Kenji Hasegawa\altaffilmark{1}.
        }
    \altaffiltext{1}{Division of Particle and Astrophysical Science, Graduate School of Science, Nagoya University, Aichi 464-8602, Japan}
    \shortauthors{M. I. N. Kobayashi et al.}
    \email{\myemail}

   


\begin{abstract}
Recent radio observations show that the giant molecular cloud (GMC) mass functions
noticeably vary
across galactic disks. High-resolution magnetohydrodynamics
simulations show that multiple episodes of compression are required
for creating a molecular cloud in the magnetized interstellar medium. In this article, 
we formulate the evolution equation for the GMC mass function to reproduce the observed profiles, 
for which multiple compression are driven by the network of expanding shells 
due to H{\sc ii} regions and supernova remnants.
We introduce the cloud-cloud collision (CCC) terms in the evolution equation in contrast to the
previous work \citep[]{Inutsuka2015}. The computed time evolution suggests that 
the GMC mass function slope is governed by the ratio
of GMC formation timescale to its dispersal timescale, and that 
the CCC effect is limited only in the massive-end of the mass function.
In addition, we identify a gas {\it resurrection}\ channel that allows 
the gas dispersed by massive stars to regenerate GMC populations or to accrete
onto the pre-existing GMCs. 
Our results show that 
almost all of the dispersed gas contribute to the mass growth of pre-existing GMCs in arm regions 
whereas less than 60\% in inter-arm regions. 
Our results also predict that GMC mass functions have a single power-law exponent 
in the mass range $<10^{5.5} \msun$ (where $\msun$ represents the solar mass),
which is well characterized by GMC self-growth and dispersal timescales.
Measurement of the GMC mass function slope 
provides a powerful method 
to constrain those GMC timescales and the gas resurrecting factor
in various environment across galactic disks.
\end{abstract}






\section{INTRODUCTION}
\label{sec:introduction}

Giant molecular clouds (hereafter GMCs) are massive and cold molecular gas reservoir for star formation
($\gtrsim 10^4 \msun$ and $10 - 100$ pc; see \citealt[][]{Williams2000,Kennicutt2012}) 
and are thus essential to study star formation 
and subsequent galaxy evolution. 
Especially, GMC properties can play a pivotal role
in governing star formation and eventually the evolution of galaxies;
GMC distribution, density, mass function, \etc\ differ between
galactic environment (\eg bulge/arm/inter-arm regions, galactic star formation rates, 
galaxy morphologies, 
redshifts, \etc; see \citealt[]{Colombo2014a, Utomo2015, Tosaki2016};
\cf \citealt[]{Tacconi2010} for giant molecular clumps at a higher redshift).
Therefore, complete understanding of galaxy evolution requires proper model of
the GMC formation and subsequent star formation in different galactic environment.
To investigate the life cycle of GMCs  
(\eg, their formation process, lifetime, star formation
within GMCs, \etc), various physics need to be understood
(\eg self gravity, magnetic fields, cosmic rays, and so on).
Recent theories and observations in the Galaxy advance our knowledge
that the filamentary structure of dense molecular gas is important
to understand the star formation within individual GMCs
\citep[]{Inutsuka2001,Andre2010,Andre2011,Roy2015}.
The study of the GMC evolution is expected to be further proceeded
based on this filamentary paradigm.
\par

Provided that star formation within GMCs is actively investigated (such as for the filament paradigm mentioned above), 
the GMC mass function (GMCMF) becomes the key to understand the star formation on galactic scales
because 
the star formation rate diversity across galactic disks
may simply originate in the diversity of GMC populations on galactic scales
if the star formation efficiency in individual GMCs is universal, as
in the local star forming clouds (\cf, Izumi et al.\ in prep).
The GMCMF is actively studied in the solar neighborhood 
(\eg, \citealt{Yonekura1997, Williams1997, Kramer1998}; also see the review by \citealt{Heyer2015}).
On the other hand,
such statistical study was
difficult to conduct in external galaxies because of two reasons:
the difficulties to identify individual GMCs in distant galaxies,
and the enormous exposure time required to detect weak molecular line emissions from low mass GMCs
$\lesssim 10^6 \msun$.

Recently, however, large radio observations with exquisite resolution started to map 
the whole disks of nearby galaxies in detail 
and to shed lights on GMC's statistical properties on galactic scales
\citep[\eg][]{Engargiola2003,Rosolowsky2003,Rosolowsky2007, Koda2009, Koda2011, Koda2012, Schinnerer2013, Colombo2014a, Colombo2014b}.
Especially, the Plateau de Bure Interferometer(PdBI) Arcsecond Whirlpool Survey (PAWS) program \citep[]{Schinnerer2013} 
demonstrate that the GMCMF varies on galactic scales:
shallow slopes in arm and central bar regions whereas steep slopes in inter-arm regions
\citep[][]{Colombo2014a}, which indicates that massive GMCs are less likely to
be formed in inter-arm regions. 
There are also some reports on the GMCMF variation along the galactocentric radii
(\eg, in the Galaxy \citep[]{Rice2016} and in M33 \citep{Rosolowsky2003, Rosolowsky2007}; 
\cf the outer Galaxy \citep{Heyer2001}, LMC\citep{Fukui2001}, 
and M83 (Hirota et al.\ in prep.))
Presumably, this type of statistical GMC studies will proceed
further 
down to both smaller mass scales $\lesssim 10^4 \msun$ and smaller spatial scales 
thanks to ongoing latest observations (\eg, in
Atacama Large Millimeter/Submillimeter Array (ALMA); \cf, \citealt{Tosaki2016}).
\par

On the theoretical side, supersonic shock compression is one of the key processes
that forms molecular clouds. The interstellar medium (ISM) has
thermally bistable atomic phases where radiative cooling balance
photo-electric heating and partially cosmic ray heating
\citep[][]{Field1969,Wolfire1995,Wolfire2003}; one phase is warm neutral medium (WNM), 
which is a diffuse H{\sc i} gas with the number density $\sim 0.1 \, \mathrm{cm^{-3}}$ 
and the temperature $\sim 6000$ K, 
and the other phase is cold neutral medium (CNM), which is an H{\sc i} cloud with the number density 
$100 \, \mathrm{cm^{-3}}$ and the temperature $\sim 100$ K. 
Supersonic shock causes the transition between these two phases
\citep[\eg,][]{Hennebelle1999,Hennebelle2000,Koyama2000}.

Previous studies with hydrodynamics simulations 
investigated the CNM formation due to the thermal instability in between two colliding WNM flows
as a precursor of molecular clouds 
\citep[\eg][]{Walder1998a,Walder1998b,Koyama2002,Audit2005,Audit2008,Heitsch2005,Heitsch2006b,vazquezsemadeni2006,Hennebelle2007a,Hennebelle2007b}.
However, recent magnetohydrodynamics simulations reveal that the molecular cloud formation is significantly retarded
in the magnetized ISM \citep[\eg][]{Inoue2008,Inoue2009,Heitsch2009,Inoue2012,Kortgen2015, Valdivia2016}.
\citet[]{Inoue2008, Inoue2009, Inoue2012} demonstrate that such prevention occurs unless
supersonic shock propagates along the magnetic fields.
In the ISM however, the shocks may come from various directions
and do not necessarily propagate along the magnetic filed.
Therefore, it is expected that the shock propagation
along the magnetic field occurs on average once out of 
a few 10 times shocks. This suggests that 
multiple episodes of supersonic WNM compression is essential for 
successful molecular cloud formation.
\par

\citet[][]{Kwan1979,Scoville1979,Tomisaka1986} studied the models
of GMC growth,
in which the coagulation due to cloud-cloud collision (hereafter CCC) alone 
governs GMC growth, and the authors do not take into account H{\sc i} cloud accumulation 
expected with magnetic fields.
Therefore, the resultant GMCMF slope may be 
determined merely by the dependence of the CCC rate on GMC mass,
so that they do not necessarily correspond to the observed ones (see Section~\ref{subsec:eq_or_noneq}).

Recently, \citet[]{Inutsuka2015} have proposed a new scenario of GMC formation and 
evolution on galactic scales, which is driven by a network of expanding H{\sc ii} regions and supernovae. 
\citet[][]{Inutsuka2015} formulate a continuity equation to describe
GMC self-growth through multiple episodes of WNM compression
and GMC self-dispersal by 
massive stars that are born in those GMCs.
Their formulation suggests that 
the observed GMCMF slope variation may originate in the variation of GMC self-growth/dispersal timescales
(see also Sections~\ref{subsec:inutsuka2015} and~\ref{subsec:case4_5}). 
\par

The bubble paradigm proposed by \citet[]{Inutsuka2015} inherently include CCC
as the collision between GMCs on the surface of neighboring expanding shells.
The authors mentioned the possible CCC contribution to the GMCMF evolution.
However, their formulation does not include CCC\@.
There are claims that CCC possibly drives the most of massive star formation within the Galaxy 
\citep[\cf][]{Tan2000,Nakamura2012,Fukui2014,Torii2015,Fukui2015b,Fukui2016}.
Therefore, the GMCMF time evolution may also be modified by CCC\@.
In addition, \citet[]{Inutsuka2015} do not consider how the dispersed gas are 
recycled into the ISM, pre-existing GMCs, and newer generation of GMCs.
\par

To evaluate the combining contribution both from multiple episodes of supersonic compression and CCC,
we, in this article, first evaluate the supersonic compression as in 
\citet[]{Inutsuka2015} 
and additionally analyze the CCC effect. Based on this formulation, 
we compute the time evolution of the GMCMF and compare with observations,
which indicate the possibility for future large radio surveys to put unique constraints 
on relevant GMC timescales on galactic scales.
We also additionally introduce a gas resurrection channel 
and suggest the importance of gas resurrecting processes regulating
the GMC evolution.

This article is organized as follows.
In Section~\ref{sec:gmc}, we review recent radio observational studies and \citet[]{Inutsuka2015} model
in more detail to endorse our formulation of evolution equation, which is presented in Section~\ref{sec:coagulationeq}.
We will compute the time-integration of the evolution equation and present the GMCMF time evolution
without CCC and with CCC respectively (Section~\ref{sec:results1}). 
In Section~\ref{sec:results2}, we will also further focus on the fate of dispersed gas 
to (1) discover the relation between the GMCMF slope and the amount of resurrecting gas
and (2) insist the importance of gas resurrecting processes.
In Section~\ref{subsec:discussion}, we (1) give the possible explanation that reconciles our 
results and the CCC importance indicated by recent radio observations
and (2) explore the other possible mechanisms to reproduce the observed GMCMF slopes.
Finally, we summarize our study in Section~\ref{sec:summary}.
The further explanation of our modeling is reported in Appendices.

\section{GMC: observations and modeling}
\label{sec:gmc}
To add detailed information to Section~\ref{sec:introduction},
we summarize two previous radio surveys (PAWS and Nobeyama Radio Observatory (NRO))
and review the GMC formation scenario which we are going to be based on 
extensively throughout this article.

\subsection{Observed mass functions}
\label{subsec:observations}
The PAWS program was conducted by using PdBI
and IRAM 30 m telescope and observed the disk of the Whirlpool Galaxy M51
in $^{12}$CO(1-0) line over the 200 hours integration. 
The survey reveals the cold gas kinematics with the $\sim 40$pc spatial resolution 
and detects objects whose mass are $\geq 1.2 \times 10^5 \msun$ at the $5\sigma$ level 
\citep[][]{Pety2013}. The PAWS team subdivide M51 into seven regions 
to analyze the GMC properties in various galactic environment
(see Figure 2 in \citealt[][]{Colombo2014a}).
One of the highlighting results is the GMCMF variation; 
bar and arm regions typically have
shallower slopes (\eg, $-1.33$ in the ``nuclear'' bar region and $-1.75$ in 
the ``density-wave spiral arm'' region) whereas
inter-arm regions (and an arm region at M51 outskirts) typically have
steeper slopes (\eg, $-2.55$ in the ``downstream'' region).
Here the slope means $-\alpha$ in $n_{\rm cl} \propto m^{-\alpha}$
where $n_{\rm cl}$ is the differential number density of GMC 
and $m$ represents the GMC mass.
In this article, we aim to reproduce such variation
based on GMC scale physics.

\citet[]{Rosolowsky2007} observed the CO(1-0) line emission from Galaxy M33
using the NRO 45m telescope combined with the data
from the BIMA interferometer and the FCRAO 14m telescope. The arm/inter-arm 
comparison shows limited differences in the slopes ($-1.9$ and $-2.2$ respectively)
with difficulties in defining arms.
However, the innermost 2.1 kpc has a prominent cutoff at the massive end ($\gtrsim
4.5 \times 10^5 \, \msun$) whereas the outer regions up to 4.1 kpc do not.
Such galactocentoric radial variation is beyond our current scope but
needs to be investigated in future.

\subsection{GMC formation and evolution driven by the network of expanding shells}
\label{subsec:inutsuka2015}
As described in Section~\ref{sec:introduction}, recent multiphase magnetohydrodynamics simulations
suggest that multiple episodes of compression are necessary to form molecular clouds from the magnetized ISM\@.
According to these simulation results, \citet[][]{Inutsuka2015} proposed a new scenario of GMC formation
and evolution on galactic scales (hereafter, SI15 scenario), which is driven by the network of expanding shells.
In SI15 scenario, the expanding {\it bubbles} \ correspond to 
expanding H{\sc ii} regions and the late phase of supernova remnants,
and dense H{\sc i} shell is formed on their surface as they expand.
Molecular clouds are produced at specific regions of the ISM that experiences multiple episodes of supersonic shocks 
(\ie swept up multiple times
by different expanding shells) or where neighboring expanding shells are colliding with each other.

Based on this scenario, \citet[][]{Inutsuka2015} formulate a continuity equation that gives
the time evolution of the GMCMF\@. Their formulation is two-folds: GMC formation and self-growth
due to multiple episodes of compression and GMC self-dispersal due to star formation within
those GMCs. Their results suggest that the ratio of typical timescales for formation 
and dispersal processes determines the slope of the GMCMF 
(see Section~\ref{subsec:case4_5} for the further explanation).

They estimate the formation timescale in the following manner.
The successful molecular cloud formation is limited when a supersonic shock propagates 
in an angle less than $0.26$ radian with respect to the magnetic field.
Given that the supersonic shock arrives isotropically, the success rate is
about $2\times 0.26^2 \pi / (4\pi) \sim 0.03$ per shock.
Gas in the ISM typically experiences supersonic shocks due to supernovae
every 1 Myr \citep[\eg,][]{McKee1977}. Thus, the time interval between consecutive shocks $T_{\rm exp}$
is somewhat smaller than 1 Myr because H{\sc ii} regions also create such shocks.
Overall, the typical timescale required to produce molecular clouds from WNM is given 
as $T_{\rm exp}/0.03 \sim 10$ Myr, which we opt to use our fiducial formation timescale
in this article (see Section~\ref{subsec:growth}).

\section{COAGULATION EQUATION}
\label{sec:coagulationeq}
Based on SI15 scenario described in Section~\ref{subsec:inutsuka2015},
we now introduce our formulation including the CCC term
to compute the time evolution of the GMCMF.
The evolution of differential number density of GMCs with mass $m$, $n_{\rm cl}$, is given as:
\begin{equation}
\begin{aligned}
    \frac{\partial n_{\rm cl}}{\partial t} +& \frac{\partial}{\partial m} 
    \left( n_{\rm cl} \left(\frac{{\rm d}m}{{\rm d}t}\right)_{\rm self} \right) \\
    =& -\frac{n_{\rm cl}}{\mytd} \\
    &+ \frac{1}{2} \int_0^\infty \int_0^\infty  K(m_1, m_2) n_{{\rm cl},1} n_{{\rm cl},2} \\
    &~~~~~~~~~~~~~~~~~~~ \times \delta(m-m_1-m_2) {\rm d}m_1 {\rm d}m_2  \\
    &- \int_0^\infty K(m, m_2) n_{\rm cl} n_{{\rm cl},2} {\rm d}m_2 \,,
    \label{eq:coageq}
\end{aligned}
\end{equation}
where $({\rm d}m/{\rm d}t)_{\rm self}$ is the mass gain rate of GMCs due to 
their self-growth, $\mytd$ is the timescale of GMC self-dispersal,
$n_{{\rm cl},1}$ and $n_{{\rm cl},2}$ are the differential number densities
of GMCs with mass $m_1$ and $m_2$ respectively, $K(m_1, m_2)$ is the 
kernel function on the collision between GMCs with $m_1$ and $m_2$, and $\delta$ is the Dirac delta function.

For each term in Equation~(\ref{eq:coageq}), we give 
ample description in the following subsections. We also discuss
the detailed variation from this fiducial formulation 
in Appendix~\ref{sec:variouseffects}.

\subsection{Self-growth term}
\label{subsec:growth}
The second term on the left hand side of Equation~(\ref{eq:coageq}) represents
the GMC self-growth. 
This term is a flux term in the conservation law. 
The ordinary continuity equation in fluid mechanics is 
a simple example of an analogous conservation law, 
which considers the mass conservation in configuration space, 
whereas our equation describes GMC number conservation in GMC mass space.

We consider that $({\rm d}m/{\rm d}t)_{\rm self}$, the GMC self-growth speed, is determined by
the multiple H{\sc i} cloud compression,
which depends on the shape of GMCs.
Observations suggest that the GMC column density does not vary much between GMCs 
\citep[\eg typically a few times $10^{22} \, {\rm cm^{-2}}$: \cf][]{Onishi1999, Tachihara2000}; 
therefore, we can assume that GMCs have rather pancake shape than perfect spherical structure,
which suggests that the GMCs' surface area is roughly proportional to their mass. 
In addition, the amount of H{\sc i} cloud accumulated onto pre-existing GMCs 
through the multiple episodes of compression
is presumably proportional to the GMC's surface area. 
Altogether, $({\rm d}m/{\rm d}t)_{\rm self}$ should be proportional to GMC mass divided by
a typical self-growth timescale that is independent of mass:
\begin{equation}
    \left(\frac{{\rm d}m}{{\rm d}t}\right)_{\rm self} = \frac{m}{\mytsg} \,.
    \label{eq:selfsg}
\end{equation}
where $\mytsg$ is a typical timescale for the GMC self-growth. 

In SI15 scenario (the bubble paradigm), GMCs are formed via multi-compressional processes, which also cause the 
self-growth of GMCs. Therefore, in our calculation, we adopt that $\mytsg$ is comparable 
to the typical GMC formation timescale, $\mytf$, which is estimated as a few 10 Myrs 
as discussed in Section~\ref{subsec:inutsuka2015} \citep[\cf][]{Inoue2012, Inutsuka2015}.
The resultant $({\rm d}m/{\rm d}t)_{\rm self}$ becomes:
\begin{equation}
    \left(\frac{{\rm d}m}{{\rm d}t}\right)_{\rm self} = \frac{m}{\mytf} \,.
    \label{eq:selftf}
\end{equation}

Equation~(\ref{eq:selftf}) with a constant $\mytf=10$ Myr indicates that GMCs grow exponentially in mass.
Given the minimum mass for GMCs ($\mymmin$) are $\sim 10^4 \msun$ (see Appendix~\ref{subsec:minimum_gmc}),
GMCs require at least 100 Myr for the exponential growth up to $10^{6.5} \msun$ (see also Appendix~\ref{subsec:init}).
With additional $14$ Myr required for destroying GMCs due to star formation (see Section~\ref{subsec:dispersal}),
we expect that observed massive GMCs $\sim10^{6.5} \msun$ typically 
have their ages $\gtrsim 114$ Myr.
$114$ Myr is almost comparable or larger than a typical timescale for the half galactic rotation $\sim 100$ Myr.
Therefore, this $114$ Myr indicates that massive GMCs $\sim10^{6.5} \msun$ in inter-arm regions are not directly 
formed ``in-situ'' in inter-arm environment; they may be remnants that were originally born in arm environment
and survived the destructing processes (\eg, by stellar feedback and galactic shear).
Modeling such transition from arm environment into inter-arm environment should be 
investigated further to study the observed spur features extended from spiral arms
(\eg, \citealt[][]{Corder2008, Schinnerer2017})
and flocculent spiral arms (\eg, in Galaxy M33).
However, we leave this for future studies and focus on
the GMCMF variation purely due to the environmental differences.
Note that $114$ Myr is the ``age'' of large GMCs $\sim10^{6.5} \msun$ but not the typical
GMC ``lifetime'' (see Section~\ref{subsec:dispersal}).

We should note here that the above formulation over-estimates the growth rate 
of very massive GMC whose mass is comparable to the mass of a shell 
swept up by an expanding bubble. 
Once the GMC mass is comparable to or larger than the typical mass of a swept-up shell, 
the growth in mass should saturates, because the dense gas that can be used to form a cloud 
is limited by the amount of total mass in the expanding shell.
Indeed, observations have revealed that GMCs exist only up to ~$10^8 \msun$ 
\citep[\eg,][]{Rosolowsky2007, Colombo2014a}. 
For modeling such gas shortage, we modify $\mytf$ by applying a growing factor 
with a truncation mass $\mtrunc$ as: 
\begin{equation}
    \mytf(m) = \mytffid\left( 1+\frac{m}{\mtrunc} \right)^\gamma \,.
    \label{eq:myratio_mdep}
\end{equation}
Here the subscript ${\rm fid}$ stands for the fiducial constant value (\ie, $\mytffid= 10 \ \mathrm{Myr}$)
and the exponent $\gamma$ determines the gas-deficient efficiency. 
The Taylor series expansion of Equation~(\ref{eq:myratio_mdep}) gives 
$\mytf(m) \simeq \mytffid (1+ \gamma m/\mtrunc)$.
Therefore, when GMCs grow up to $\mcrit \sim \mtrunc/\gamma$, $\mytf$ deviates
longer than $\mytffid$ so that the choice of $\mtrunc$ 
and $\gamma$ modify the massive end of the GMCMF\@.
Essentially, $\mcrit$ represents the typical maximum GMC mass that can be created in
SI15 scenario. In this scenario, GMCs are created from interstellar medium 
swept up by supersonic shock compression, thus it is less likely to form GMCs 
more massive than the total mass that a single supernova remnant can sweep.
The total mass initially contained within a sphere of $100$ pc radius
with H{\sc i} density $10 \, \mathrm{cm^{-3}}$ is about $7.7 \times 10^5 \msun$.
Thus $\mcrit = 7.7 \times 10^5 \msun$ and this gives $\mtrunc = 7.7 \times 10^6 \msun$ 
given that we opt to use $\gamma=10$.

The detailed modeling of $\mtrunc$ and $\gamma$ change the relative importance of
GMC self-growth/dispersal and CCC\@.
In addition, rapid star formation triggered by CCC also needs to be properly modeled
if CCC becomes effective (see Section~\ref{subsec:sf_ccc}).
However, these details would not largely impact 
if we focus on the GMCMF slope
(see Section~\ref{subsec:case1_2}).
Therefore, we will reserve the detailed investigation for future works.
\par
Note that the second term on the left hand side of 
Equation~(\ref{eq:coageq}) has its boundary condition
at $m=\mymmin = 10^4 \msun$. The flux at this boundary in our formulation corresponds to the minimum-mass 
GMC production rate. In later sections, we will explain that this rate differs between setups 
(see the first paragraph
in Section~\ref{sec:results1} and the second paragraph in Section~\ref{sec:results2}).

\begin{table*}
    \caption{Studied setups}
    \centering{
        \begin{tabular}{cccccc}
            \hline
            \hline
            \input{cases.table}
            \hline
            \hline
        \end{tabular}
    }\par
    \bigskip
    \textbf{Note.} 8 cases that we studied. Individual parameters are described in Section~\ref{sec:coagulationeq}. 
              Each column represents as follows; $\mytffid$ shows the formation timescale. 
              $\mytd$ gives the dispersal timescale. 
              Cloud-Cloud Collisions indicate whether we include the CCC terms (yes) or not (no). 
              $\myrec$ represents the resurrecting factor
              introduced in Section~\ref{subsec:recycle} (\ie the fractional mass out of total dispersed gas 
              that is consumed to 
              form newer generation of GMCs), where ``support'' means that we artificially 
              keep the population of minimum-mass GMCs fixed.
              Figures indicate the corresponding figures in this article.
    \label{table:cases}
\end{table*}

\subsection{Dispersal term}
\label{subsec:dispersal}
The first term on the right hand side gives the GMC self-dispersal rate.
Here, self-dispersal means that massive stars born within GMCs
destroy their parental GMCs by any means (ionization, dissociation, heating, blowing-out, etc.).
The characteristic self-dispersal timescale, $\mytd$, is given as:
\begin{equation}
    \mytd = \mytsb + \mytsd \,,
\end{equation}
where $\mytsb$ is the typical timescale for star formation after GMC birth
and $\mytsd$ is the typical timescale for GMC destruction 
once stars become main-sequence stars.
We assume the typical star formation timescale within GMCs is $\sim \mytffid$
and thus employ $\mytsb = 10$ Myr.
$\mytsd$ can be estimated by line-radiation magnetohydrodynamics simulations. 
\citet[][]{Inutsuka2015} reported one of such simulation
results, which updated the work in \citet[][]{Hosokawa2006b} by including 
magnetic fields. Their results indicate that, once a massive star with mass $>30 \msun$ is formed in a GMC,
the star destroys more than $10^5 \msun$ surrounding molecular hydrogen within 4 Myr.
Therefore, typical $\mytd$ is $10 + 4 = 14$ Myr.

$\mytd$ can be understood as the typical GMC ``lifetime'' 
averaged over all the populations
because $\mytd$ is the typical timescale within which GMCs disperse in the system.
This is consistent with the short lifetime indicated by observations
(\eg, in LMC \citep[][]{Kawamura2009} and in M51 \citep[][]{Meidt2015}).

$\mytsd$ is basically independent of GMC mass because, for example, 10 times massive GMCs generate 10 times more
$>30 \msun$ stars, which in turn results in destroying 10 times more molecular gas
\citep[see][]{Inutsuka2015}.
Therefore, if $\mytsb$ is irrespective of GMC mass, $\mytd$ does not
depend on GMC mass. Thus, we opt to use $\mytd=14$ Myr throughout the current article.
Note that this argument of mass independence is valid only if parental GMCs are $>10^5 \msun$
because this is the minimum mass required to generate stars $>30 \msun$ if we adopt
Salpeter initial stellar mass function \citep[]{Salpeter1955}.
In case of GMCs $<10^5 \msun$, the self-dispersal may be less effective \citep[see][]{Inutsuka2015}
and more careful treatment is desired. In addition, 
$\mytsb$ can be one order of magnitude shorter if 
CCC triggers rapid formation of massive stars
\citep[\cf][]{Fukui2016}
(see also Section~\ref{subsec:sf_ccc}).
For simplicity, however, we do not consider any $\mytsb$ and $\mytsd$ variation in the current article
(see also Section~\ref{subsec:sf_ccc}).

\subsection{Cloud-Cloud Collision terms}
\label{subsec:ccc_term}
The second and third terms on the right hand side represent the collisional coagulation
between GMCs (\ie the CCC process).
The second term on the right hand side produces a
GMC with mass $m (=m_1 + m_2)$ through CCC between GMCs with their mass $m_1$ and $m_2$.
Similarly, the last term on the right hand side creates a GMC with mass $m+m_2$ 
through CCC between GMCs with their mass $m$ and $m_2$,
so that the negative sign indicates that this collision decreases the number density of GMCs with mass $m$.
$K(m_1,m_2)$ is the product of the total GMC collisional cross section between GMCs 
with mass $m_1$ and $m_2$, $\sigma_{{\rm col} \, 1,2}^{}$,
and the relative velocity between GMCs, $V_{\rm rel}$:
\begin{equation}
    K(m_1,m_2) = \sigma_{{\rm col} \, 1,2}^{} \, V_{\rm rel} 
    = c_{\rm col}\frac{m_1 + m_2}{\Sigma_{\rm mol}^{}} V_{\rm rel,0} \,.
    \label{eq:kernel}
\end{equation}
Here, $c_{\rm col}$ is a correction parameter, $\Sigma_{\rm mol}^{}$ is a GMC column density,
and $V_{\rm rel,0}$ is the fiducial relative velocity between GMCs.
$c_{\rm col}$ reflects various effects (\eg, geometrical structure, gravitational focusing, 
relative velocity variation) 
and is expected to be on the order of unity (see discussions in Appendix~\ref{sec:variouseffects}).
For simplicity, we use $c_{\rm col}=1$ throughout this article.
For the column density and the relative velocity, we opt to employ an observed constant value 
$\Sigma_{\rm mol} = 2 \times 10^{22} \mu m_{\rm H}^{} \, \mathrm{cm^{-2}}$
(see Section~\ref{subsec:growth}) and a typical expanding speed of ionization-dissociation front 
$V_{\rm rel,0} = 10 \, \mathrm{km \, s^{-1}}$ (see Appendix~\ref{subsubsec:relvel}) respectively,
where $\mu$ is the mean molecular weight and $m_{\rm H}^{}$ is the atomic hydrogen mass.

Our calculation focuses only on collisional coagulation and ignores fragmentation at all.
This assumption is based on the observational fact that cloud-to-cloud velocity dispersion 
($\sim 10 \, \mathrm{km \, s^{-1}}$: \eg, \citealt[][]{stark2005}) does not largely exceed 
the sound velocity ($\sim 10 \, \mathrm{km \, s^{-1}}$) 
of inter-cloud medium (\ie, WNM).
However, the fragmentation may have a severe impact under the CCC-dominated phase
(\ie, the GMC number density is high)
such as in the Galactic Center \citep[]{Tsuboi2015}.
This is beyond our current scope and we will report 
in our forthcoming paper.

Note that $\mytd$ in principle should be the ensemble average of all the destructive processes 
including, for example, shear \citep[\cf,][]{Koda2009}.
However, shear may be sub-dominant in the disk regions typically $>5$ kpc \citep[\cf,][]{Dobbs2013}.
For simplicity, we assume that the inclusion of these processes would not vary $\mytd$ significantly.

\section{Results 1: Slope of Giant Molecular Cloud Mass Function}
\label{sec:results1}
We perform the time integration of Equation~(\ref{eq:coageq})
to determine what controls the observed GMCMF\@. Table~\ref{table:cases} lists the
parameters we use; especially Cases 1 to 4 correspond to the analysis in this section.
Throughout this section, we assume that minimum-mass GMCs are always
continuously created so that the rate producing new-born minimum-mass GMCs
compensates the decrease in the population of minimum-mass GMCs due to self-growth, self-dispersal, and CCC\@.
This treatment makes the number density of minimum-mass GMCs 
always constant. 
This steadiness of minimum-mass GMCs is also presumed from
relatively constant star formation activity observed in nearby galaxies 
\citep[]{Madau2014}.
Without any sufficient creation, minimum-mass GMCs are 
exhausted and the observed GMCMF cannot be reproduced.
In Section~\ref{sec:results2}, we will discuss
possible variation in the rate of producing new-born minimum-mass GMCs
by dispersed gas ``resurrection'' (\ie, regeneration of GMC populations
from dispersed gas produced by massive stars).

\begin{figure}\centering{
\includegraphics[width=0.80\columnwidth,keepaspectratio]{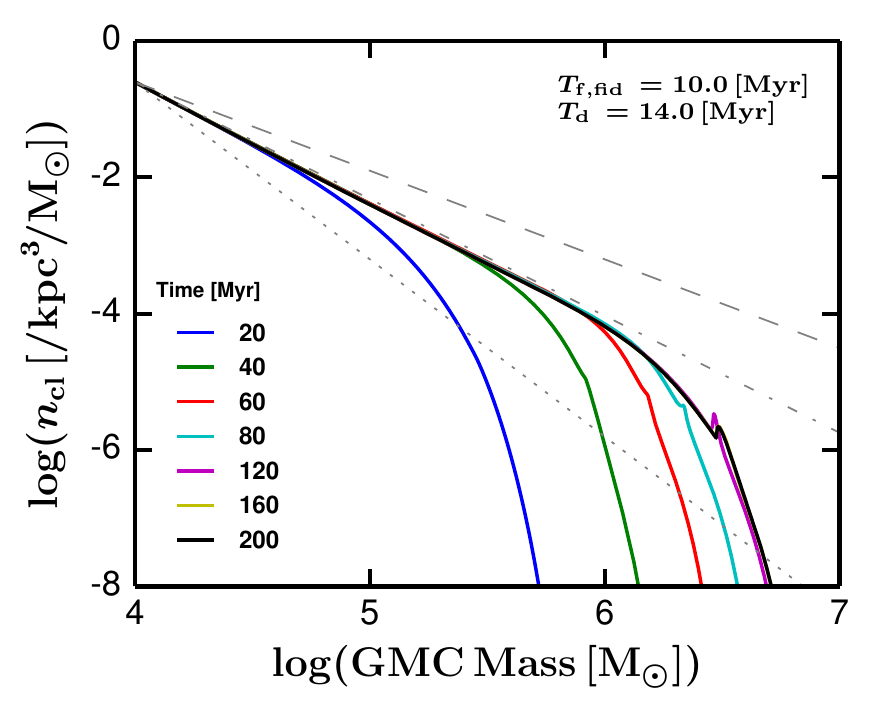}}
\caption{Case 1: Differential number density $n_{\rm cl}$ as a function of GMC mass,
based on the fiducial calculation with CCC\@. The color corresponds to the time evolution. 
Note that the yellow line corresponds to $160$ Myr is almost fully
overlapped by the black line for $200$ Myr, thus it is not visible on this plot.
As a reference, we plot three gray straight lines; 
dot-dashed line fits the computed GMCMF slope ($\sim -1.7$), 
dashed line corresponds to the observed shallow slope whereas the dotted line shows the observed steep slope. 
The calculated GMCMF fits into the middle of the observed slope range.}
\label{fig:all_support}
\end{figure}

\begin{figure}\centering{
\includegraphics[width=0.80\columnwidth,keepaspectratio]{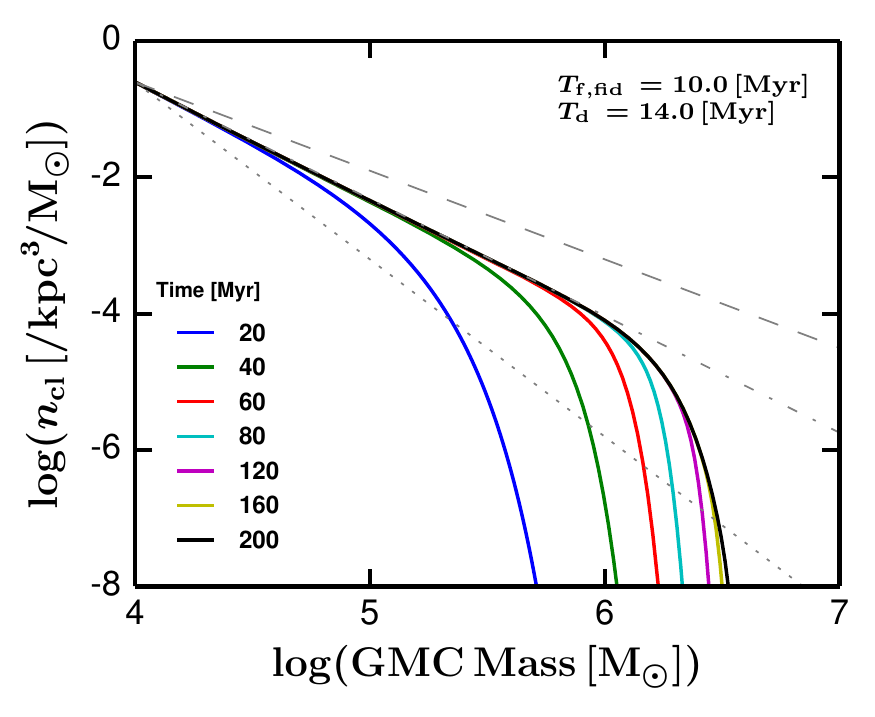}}
\caption{Case 2: Differential number density $n_{\rm cl}$ as a function of GMC mass,
based on the fiducial calculation without CCC\@. The color corresponds to the time evolution. 
As a reference, we plot three gray straight lines; dot-dashed line fits the computed GMCMF slope ($\sim -1.7$), 
dashed line corresponds to the observed shallow slope whereas the dotted line shows the observed steep slope. 
The calculated GMCMF fits into the middle of the observed slope range.
The similarity with Figure~\ref{fig:all_support} indicates that CCC does not impact on the GMCMF slope significantly
but modify the massive end.}
\label{fig:self_support}
\end{figure}

We set the timestep width as 0.1\% of the shortest timescale in each time step.
We use the logarithmically spaced mass grid where the $i+1$-th mass is larger by a factor 1.0125 than the 
$i$-th mass.
The initial GMCMF has a delta-function like mass distribution where
only minimum-mass GMCs exist, by which we can highlight
how fast the GMCMF evolves.
All the figures presented here show the GMCMF up to the mass where the cumulative number
of GMCs $>1$, except that the GMCMFs in Figs.~\ref{fig:all_support} and~\ref{fig:self_support} 
are shown beyond that mass to show the CCC effect clearly,
and also except that the GMCMF in Fig.~\ref{fig:fid_rec1} is shown below that mass 
because GMCs more massive than the observed ones are created 
due to our rather artificial choice of an excessively CCC-dominated condition
and we would like to focus on the slopes in the observed mass range.
Note that we turn off the CCC calculation between certain GMC pairs when at least one of GMCs in the pair has its cumulative number less than 1 
because such GMCs are less likely to exist in real galaxies so that this type of CCC is also expected to be rare.
The spikes and kinks that appear in the massive end (especially in Figs.~\ref{fig:all_support} and~\ref{fig:all_longtf})
are due to the numerical effect by this CCC turn-off procedure so that actual GMCMFs in the Universe would 
be smoother.
The results of PAWS survey \citep[\cf,][]{Colombo2014a} are used in the plots
for comparing the GMCMF slopes between our calculation and observations.

\begin{figure}\centering{
\includegraphics[width=0.80\columnwidth,keepaspectratio]{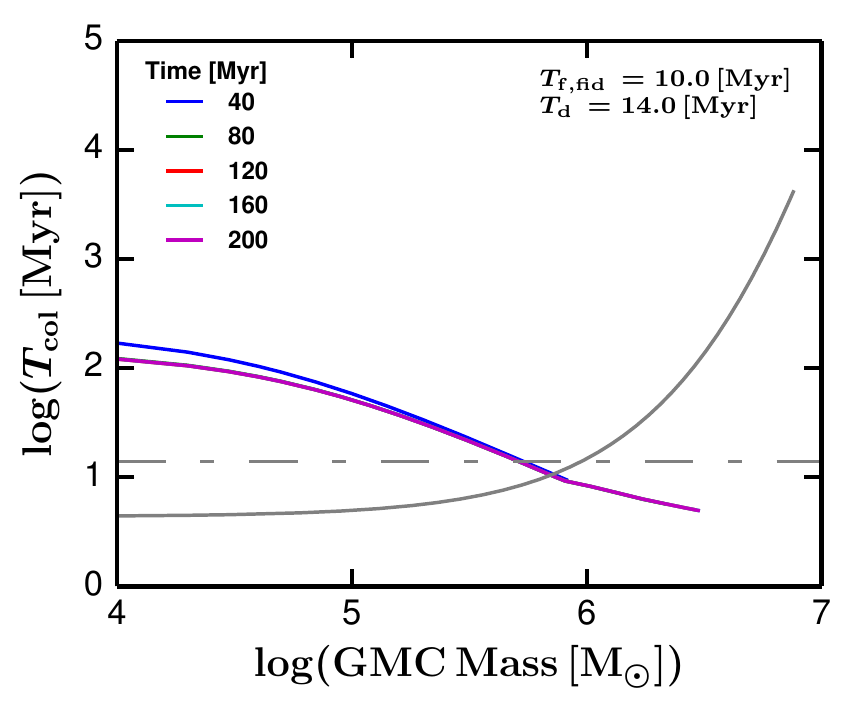}}
\caption{The three timescales as a function of GMC mass in Case 1. 
The gray solid line represents $\mytf$, the gray horizontal dot-dashed line corresponds to $\mytd$, 
and the colored lines show the time evolution of $\mytcol$.
Note that the three colored lines (\ie, the green line for $80$ Myr,
the red line for $120$ Myr, and the cyan line for $160$ Myr) are overlapped by
the purple line for $200$ Myr so that they are not visible on this plot.
The figure indicates that GMC self-growth dominates in low-mass regime, but CCC deforms the GMCMF 
at its high-mass end where $\mytcol$ is one order of magnitude smaller than self-growth timescale.
Note that the gray line of $\mytf$ shown here has a factor difference from original $\mytf$ 
defined as Equation~(\ref{eq:myratio_mdep}) (see Appendix~\ref{sec:tflow}).}
\label{fig:tcol_all_support}
\end{figure}

\subsection{CCC contribution to the GMCMF slopes and massive ends}
\label{subsec:case1_2}
Figures~\ref{fig:all_support} and~\ref{fig:self_support} show our fiducial calculations
(\ie, $\mytffid = 10$ Myr, $\mytd = 14$ Myr), which correspond to Case1 and Case2
respectively. The only difference between these two cases is whether
the calculation includes the CCC terms (Figure~\ref{fig:all_support}) 
or not (Figure~\ref{fig:self_support}). Both computed GMCMFs demonstrate
almost the same slope $-\alpha \sim-1.7$ (where $n_{\rm cl} \propto m^{-\alpha}$), 
which successfully fits into the middle of the observed range.
This result indicates that CCC does not impact the GMCMF slope significantly whereas
the shape of the massive end is modified by the relative importance 
between self-growth/dispersal and CCC\@. 

To examine the relative contribution of CCC, we compute the collision timescale, $\mytcol$, as:
\begin{equation}
    \mytcol = \frac{n_{\rm cl}(m)}{\left( \partial n_{\rm cl} / \partial t \right)_{\rm CCC}} \,,
    \label{eq:tcol}
\end{equation}
where the denominator is the time evolution due to CCC:
\begin{equation}
\begin{aligned}
\left( \frac{\partial n_{\rm cl}}{\partial t} \right)_{\rm CCC} = 
    & \frac{1}{2} \int_0^\infty \int_0^\infty  K(m_1, m_2) n_{{\rm cl},1} n_{{\rm cl},2} \\
    &~~~~~~~~~~~~~~~~~~~ \times \delta(m-m_1-m_2) {\rm d}m_1 {\rm d}m_2  \\
    &- \int_0^\infty K(m, m_2) n_{\rm cl} n_{{\rm cl},2} {\rm d}m_2  \,.
    \label{eq:derivative_tcol}
\end{aligned}
\end{equation}
Figure~\ref{fig:tcol_all_support} shows the time evolution of computed $\mytcol$, 
with $\mytf$ and $\mytd$ overplotted together.
The figure indicates that the GMC mass growth is determined by GMC self-growth 
in low-mass regime where $\mytf$ ($\sim \mathcal{O}(1)$ Myr) 
is one order of magnitude shorter than $\mytcol$ ($\sim \mathcal{O}(2)$ Myr),
and is determined by CCC at the high-mass end where
$\mytcol$ ($\sim \mathcal{O}(1)$ Myr) becomes one order of magnitude smaller than $\mytf$
($\sim \mathcal{O}(2)$ Myr).
This indicates that the GMCMF slope in $m \lesssim 10^{5.5} \msun$ is well characterized by the combination of 
the GMC self-growth and dispersal. 
In the next section, we are going to focus only on the GMCMF slopes
and will discuss the massive-end behavior in Section~\ref{subsec:sf_ccc}.

\subsection{Characteristic Slope of the GMCMF}
\label{subsec:case4_5}
\begin{figure}\centering{
\includegraphics[width=0.75\columnwidth,keepaspectratio]{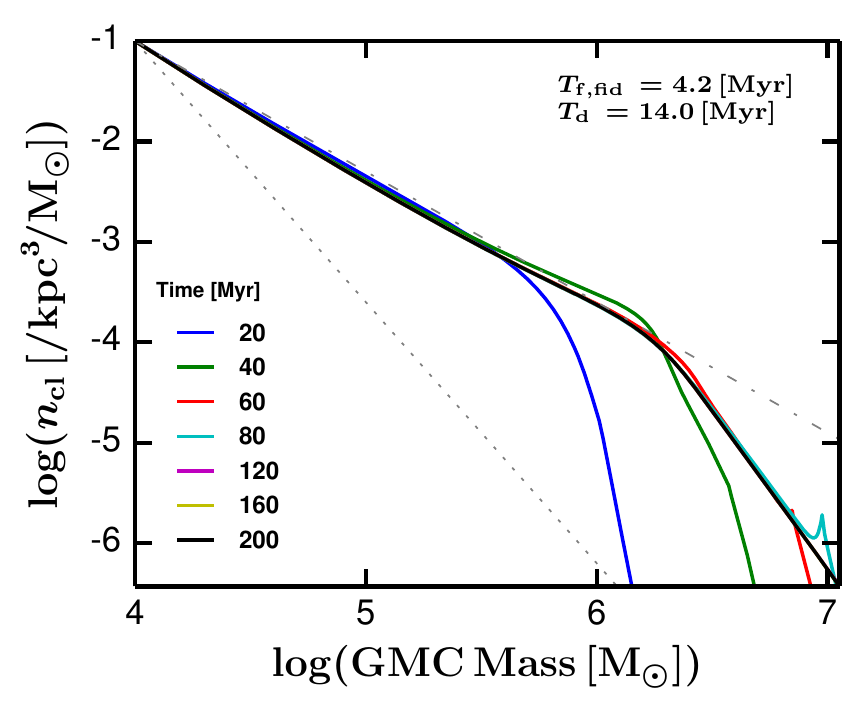}}
\caption{Case 3: Differential number density $n_{\rm cl}$ as a function of GMC mass,
based on the calculation with $\mytffid =4.2$ Myr including CCC\@. The color corresponds to the time evolution. 
Note that the two colored lines (\ie, the purple line for $120$ Myr and the yellow line for $160$ Myr)
are overlapped by the black line for $200$ Myr so that they are not visible on this plot.
As a reference, we plot two gray straight lines; dot-dashed line corresponds to the observed GMCMF slope ($\sim -1.3$)
in the arm regions, whereas the dotted line shows the observed steep slope in inter-arm regions.
The agreement between the observed slope and the computed slope suggests that the arm regions typically have a
shorter self-growth timescale, $\mytf \sim 4$ Myr. This is well characterized by the steady state solution without CCC 
shown in Equation~(\ref{eq:inutsuka_slope}). Again, the CCC effect is limited but modifies the GMCMF massive-end.}
\label{fig:all_shorttf}
\end{figure}

\begin{figure}\centering{
\includegraphics[width=0.75\columnwidth,keepaspectratio]{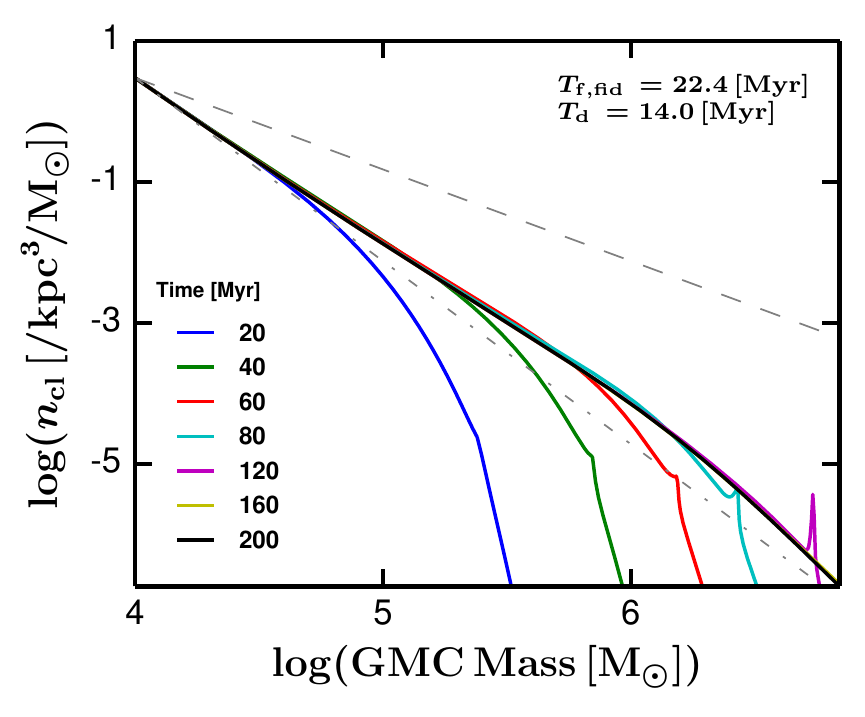}}
\caption{Case 4: Differential number density $n_{\rm cl}$ as a function of GMC mass,
based on the calculation with $\mytffid =22.4$ Myr including CCC\@. The color corresponds to the time evolution.
Note that the yellow line for $160$ Myr is overlapped by the black line for $200$ Myr, thus it
is not visible on this plot.
As a reference, we plot two gray straight lines; the dot-dashed line represents the observed steep slope ($\sim -2.6$)
in the inter-arm regions, whereas the dashed line shows the observed shallow slope in the arm regions.
Although the GMCMF shows the slope $\sim -2.2$ relatively shallower than the observed one $\sim -2.6$ 
due to the coagulation by CCC, the basic correspondence suggests that the inter-arm regions typically have a
longer self-growth timescale $\mytf \sim 22$ Myr. This is well characterized by the steady state solution without CCC 
shown in Equation~(\ref{eq:inutsuka_slope}).}
\label{fig:all_longtf}
\end{figure}

As shown in Section~\ref{subsec:case1_2}, CCC does not modify the GMCMF evolution significantly.
Especially in lower mass range (\eg, $m< 10^{5.5}\msun$) CCC does not affect the mass function, and hence
our formulation 
can be rewritten without the CCC terms. Therefore, 
the evolution of differential number density of GMCs with mass $m$, $n_{\rm cl}$, is now 
simply given as:
\begin{equation}
    \frac{\partial n_{\rm cl}}{\partial t} + \frac{\partial}{\partial m} 
    \left( n_{\rm cl} \frac{m}{\mytf} \right) 
    = -\frac{n_{\rm cl}}{T_{\rm d}} \,.
    \label{eq:inutsuka_coageq}
\end{equation}
This formulation has been already (and originally) proposed by \citet[]{Inutsuka2015}.
One can obtain the steady state solution of this differential equation for 
$m\lesssim m_{\rm crit}$ with a constant $\mytf$ as: 
\begin{equation}
    n_{\rm cl}(m) = n_0 \left( \frac{m}{\msun} \right)^{-1-\frac{\mytf}{\mytd}} \,,
    \label{eq:inutsuka_slope}
\end{equation}
where $n_0$ is the differential number density normalized at $m=1\msun$.
This predicts that GMCMFs have slopes with a single exponent, which is 
well characterized by $-1-\myratio$. 
Indeed, the computed GMCMF shows the slope consistent with this exponent $-1-\myratio$
without CCC (see Figure~\ref{fig:self_support}),
and in the lower mass ($m < 10^{5.5}\msun$) part even in the case with CCC
(see Figure~\ref{fig:all_support}).

To examine the validity of the $-1-\myratio$ prediction, we conduct the calculation again by employing various 
$\mytf$.
Figures~\ref{fig:all_shorttf} and~\ref{fig:all_longtf} are the two representative results and
both cases exhibit the slope well characterized by $-1-\myratio$.
Figure~\ref{fig:all_shorttf} employs a shorter formation timescale, $\mytffid = 4.2$ Myr, and reproduces
the observed shallow slope in arm regions. 
Similarly, Figure~\ref{fig:all_longtf}
shows the result with $\mytffid = 22.4$ Myr, which reproduces the observed steep slope 
in inter-arm regions. Therefore, the arm regions presumably have a larger number of massive stars forming
H{\sc ii} regions and supernova remnants thus experience the recurrent supersonic compression
twice more frequently than the fiducial case, 
and the inter-arm regions typically have a smaller number of massive stars
thus experience less frequent supersonic compression, 
which results in about factor two longer formation timescale than the fiducial case.
This may also explain the observed steep slope at outskirts of galaxies 
(\eg, the observed steep slope in the CO luminosity function in Galaxy M33; 
see \citealt[][]{Gratier2012}) where star formation is less active 
compared with normal disk regions. 
Note that the above argument might be modified, 
if the effect of large-scale dynamics (\eg, interaction with shock waves 
or strong shear flows) may play an important role in the destruction of GMCs 
than the stellar feedback, especially in the outskirts of galaxies with prominent spiral structures.

Note that we here assume that the $\mytd$ variation between different regions
is limited compared with the $\mytf$ variation. This is because $\mytd$ is basically independent
of GMC mass as explained in Section~\ref{subsec:dispersal} and also because observations suggest
that the star formation efficiency is almost the same throughout the galactic disks
in nearby galaxies \citep[\eg,][]{Schruba2011} and in the Galaxy (Izumi et al.\ in prep.).
However, $\mytd$ might be longer in inter-arm regions compared with arm regions
due to its less star forming activity where $\mytd$ can be up to $\sim30$ Myr
(the upper limit measured in M51 by \citealt[]{Meidt2015}),
or might be a factor longer in smaller clouds $\lesssim 10^5 \msun$
where it is invalid to apply our mass-independent assumption 
on the cloud destruction rate due to massive stars.
The significance of $\mytd$ variation should be investigated more and 
we will leave this for future work.

Our results suggest that the variation of the GMCMF slopes are governed 
by $\myratio$ diversity in different environment on galactic scales.
We predict that future large radio surveys reveal that the GMCMFs
have a single power-law exponent with which the ratio $\myratio$ may be uniquely
constrained.

\subsection{Possible Modification in the Massive-end.}
\label{subsec:sf_ccc}
The shape of the GMCMF massive-end, typically $>10^6 \msun$, is determined by the relative importance of 
GMC self-growth and CCC as shown in previous sections.
The quantitative discussion involves many uncertainties from our modeling;
for example, more detailed prescription for $\gamma$ is necessary for GMC self-growth (\ie, $\mytf$),
and it is also required to model $T_*$ variation due to the drastic star formation invoked by CCC,
which is inferred by observations in the Galaxy (\cf, \citet[]{Fukui2014};
also see Section~\ref{subsec:dichotomy}).
These are beyond our current scope and to be discussed in our forthcoming paper.

Despite these limitation, our results indicate that, if CCC becomes effective,
another structure
may appear in the massive-end. This possibly explains
the extra power-law feature observed in some regions (\eg, Material Arms) 
from PAWS survey \citep[][]{Colombo2014a}.

\section{Results 2: Fate of Dispersed Gas}
\label{sec:results2}
\subsection{Resurrecting factor}
\label{subsec:recycle}
For all the time evolutions of the GMCMF we have shown in the previous sections, 
we assume that minimum-mass molecular clouds are continuously provided 
and the dispersed gas is removed from the system and never restored into the system.

However, in reality, dispersed gas should return to the ISM and become either 
seeds of newer generation of molecular clouds or mass accreting onto the pre-existing GMCs.
To establish a complete picture of gas resurrecting processes in the ISM, we need to evaluate
the fate of such dispersed gas. This aspiration requires 
detailed numerical simulations, ideally, 
three dimensional radiation magnetohydrodynamics simulations, which are however computationally 
too expensive to conduct. Instead in this article, we quantify the amount of
dispersed gas by introducing the ``resurrecting factor'', $\myrec$, 
which is the mass fraction out of the total dispersed gas
that are transformed to newer generation of minimum-mass GMCs. 
In Section~\ref{sec:results1}, we artificially
set the production rate of minimum-mass GMCs to keep its number density constant.
Here instead, we evaluate this production rate due to gas resurrection as:
\begin{equation}
    \left. \frac{\partial \left(n_{\rm cl}m\right)}{\partial t} \right|_{\rm res} 
    = \myrec \dot{M}_{\rm total,disp} \delta(m-\mymmin) \,,
    \label{eq:fmin_res}
\end{equation}
where $\dot{M}_{\rm total,disp}$ is the total dispersed gas mass produced from the system per unit time,
$\mymmin$ is the minimum-mass of GMCs (\ie $10^4 \msun$ in this study),
and $\delta$ is the Dirac delta function.
By this definition,
$\myrec$ can be considered as the probability that 
dispersed gas becomes minimum-mass GMCs before they accrete onto the pre-existing GMCs
and thus the $1-\myrec$ fraction of dispersed gas are consumed for the self-growth of 
pre-existing GMCs if a GMCMF is in a steady state.

From now, we are going to focus only on steady state GMCMFs for simplicity.
We may assume that the GMCMFs are quasi-steady in the Galaxy and nearby massive spiral galaxies
because they presumably have already undergone active star formation phase at redshift $\sim 2$
and have relatively constant star formation activity at the present day \citep[\eg][]{Madau2014}.
Note that, although we have already proved that the CCC effect is limited, we fully solve Equation~(\ref{eq:coageq})
including the CCC terms to compute the time evolution in all the following results.
We mainly analyze Cases 5 to 8 in Table~\ref{table:cases}. 
The spikes and kinks that appear in the massive end (especially in 
Figs.~\ref{fig:fid_rec1}, ~\ref{fig:shorttf_rec5em4}, and~\ref{fig:longtf_rec038}) are 
due to the numerical effect by the CCC turn-off procedure as explained 
in the second paragraph in Section~\ref{sec:results1}.

\subsection{Not All the Dispersed Gas are Consumed for Minimum-Mass GMC Creation}
\label{subsec:case6_7}
\begin{figure}\centering{
\includegraphics[width=0.75\columnwidth,keepaspectratio]{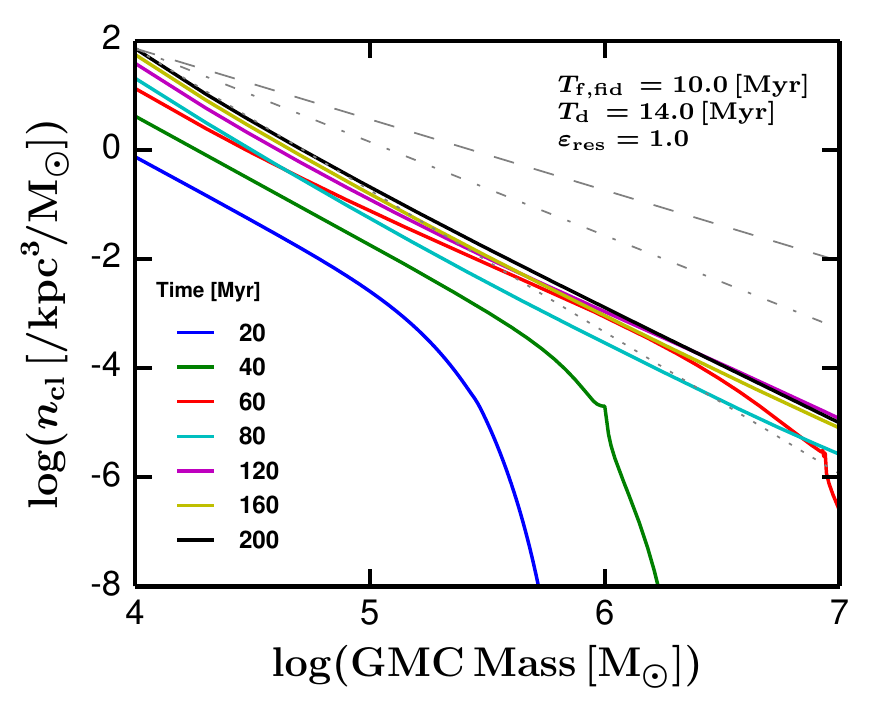}}
\caption{Case 5: Differential number density $n_{\rm cl}$ as a function of GMC mass
including CCC with the fiducial timescales ($\mytffid=10$Myr and $\mytd=14$Myr) and with $\myrec=1$.
The color corresponds to the time evolution.
As a reference, we plot three gray straight lines; the dot-dashed line is the slope $\sim -1.7$
predicted by Equation~(\ref{eq:inutsuka_slope}), the dotted line represents the observed steep slope ($\sim -2.6$)
in the inter-arm regions, whereas the dashed line shows the observed shallow slope in the arm regions.
The slope of the GMCMF is initially between $-1.7$ and $-2.6$ but 
becomes steepened and 
the total mass in the system 
keeps growing so that this does not reach any steady state (\cf, Figure~\ref{fig:fid_mtotal}). 
The kink in the massive-end at 40 Myr indicates the onset for the CCC-dominant phase.
After 40 Myr, CCC becomes dominant and massive GMCs $> 10^{7} \msun$ are continuously created,
which is not consistent with observations, either. The steepened slope
at the low-mass end appears due to overloading resurrection.}
\label{fig:fid_rec1}
\end{figure}

Here, we examine the fiducial timescales $\mytffid=10$ Myr and $\mytd=14$ Myr.
First, we employ an extreme case where $\myrec=1$; all the dispersed gas is consumed to
form minimum-mass GMCs (Case 6) and Figure~\ref{fig:fid_rec1} shows the result.
Initially the GMCMF exhibits slopes slightly steeper than $-1.7$ predicted by 
Equation~(\ref{eq:inutsuka_slope}). However the slope becomes more steepened 
in the low-mass regime and the total mass in the system keeps
increasing so that the GMCMF does not show any steady state (see Figure~\ref{fig:fid_mtotal}).
Therefore, the slopes are provisional and transitional. Indeed, 
the CCC becomes significantly effective after 60 Myr so that GMCs more massive
than observed ones are instantly created after 60 Myr,
which do not reproduce the observed GMCMFs.
\par
\begin{figure}\centering{
\includegraphics[width=0.75\columnwidth,keepaspectratio]{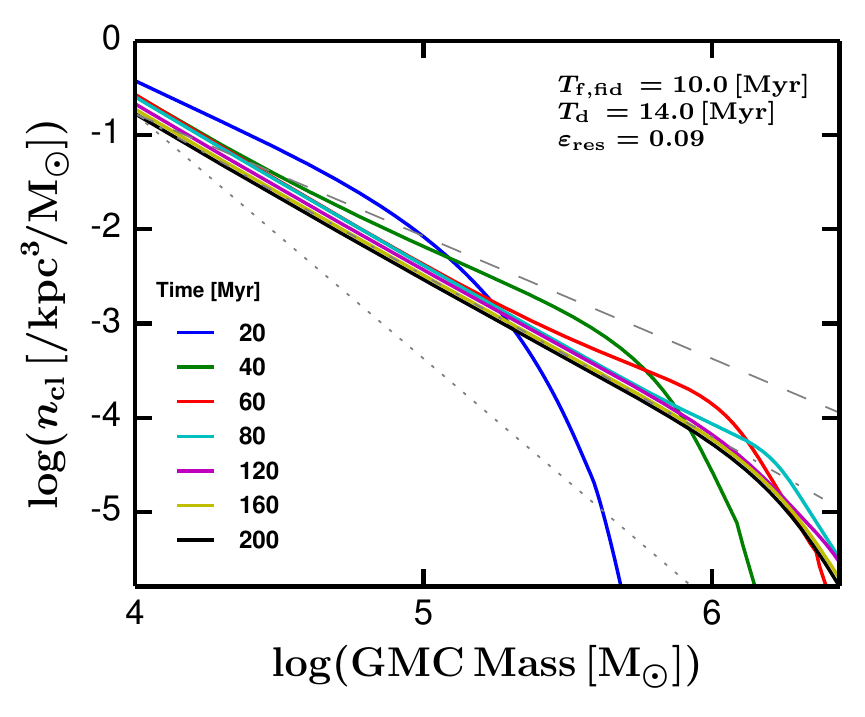}}
\caption{Case 6:
Differential number density $n_{\rm cl}$ as a function of GMC mass
including CCC with the fiducial timescales ($\mytffid=10$Myr and $\mytd=14$Myr) and with $\myrec=0.09$.
The color corresponds to the time evolution.
As a reference, we plot three gray straight lines; the dot-dashed line is the slope 
predicted by Equation~(\ref{eq:inutsuka_slope}), the dotted line represents the observed steep slope ($\sim -2.6$)
in the inter-arm regions, whereas the dashed line shows the observed shallow slope in the arm regions.
The computed GMCMF successfully reproduces the fiducial slope $\sim -1.7$ 
predicted by Equation~(\ref{eq:inutsuka_slope})
and lands on a steady state as shown in Figure~\ref{fig:fid_mtotal}.}
\label{fig:fid_rec001}
\end{figure}

\begin{figure}\centering{
\includegraphics[width=0.75\columnwidth,keepaspectratio]{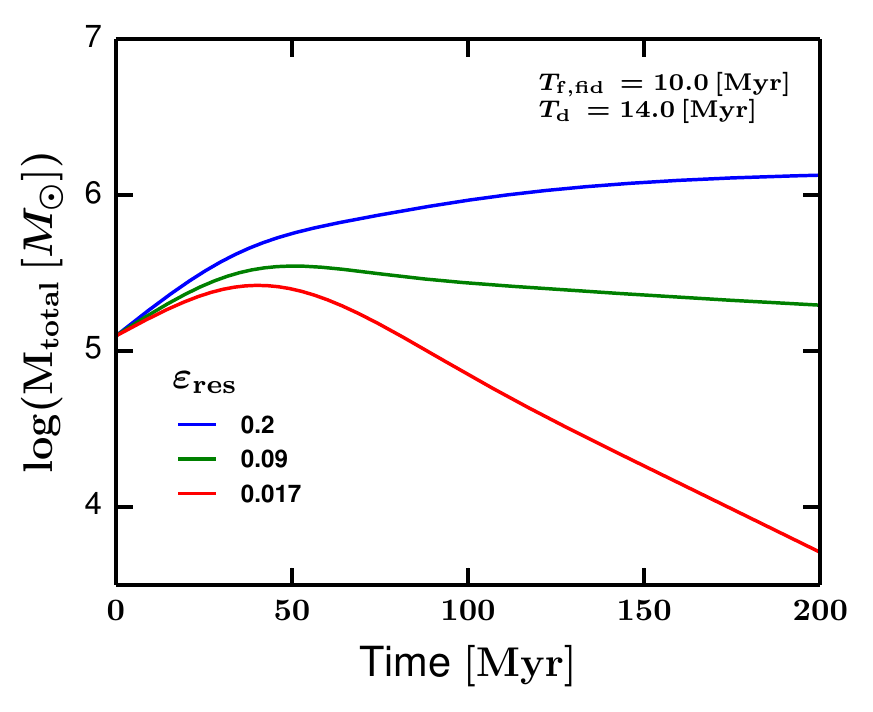}}
\caption{The total mass dependence on $\myrec$ as a function of time with the 
fiducial setup ($\mytffid=10$ Myr and $\mytd=14$Myr).
The different color represents the different $\myrec$ from $0.0017$ to $0.2$. The GMCMF with $\myrec=0.09$ 
reaches a steady state after 130 Myr elapse.
$\myrec=0.09$ reproduces the fiducial slope $-\alpha=-1.7$ as shown in Figure~\ref{fig:fid_rec001}.
Resurrecting more than $\myrec=0.09$ (\eg, $0.2$ in this figure) 
increase the total mass in the system and the GMCMF would not reach 
any steady state. Similarly, resurrecting less than $\myrec=0.09$ (\eg, $0.0017$ in this figure) 
eventually decrease the total mass in the system after 50 Myr. The steady resurrecting factor 
($\sim 9$ per cent) is predictable, which we will explain in Section~\ref{subsec:slope_and_recycle}.}
\label{fig:fid_mtotal}
\end{figure}

 
The steepened slope observed in the low-mass end in Figure~\ref{fig:fid_rec1} appears due to overloading $\myrec$,
because pre-existing GMCs cannot acquire the excessive amount of resurrecting gas faster than 
a given self-growth timescale (in this case $10$Myr). To achieve 
a steady state with a shallower slope, we need to reduce $\myrec$. 
Figure~\ref{fig:fid_rec001} shows a result
with $\myrec=0.09$ where we successfully reproduce a slope $-\alpha=-1.7$ throughout all the mass range
and achieve a steady state GMCMF\@.
To check the steadiness, we compute the total mass in the system as a function of time with various $\myrec$ 
and Figure~\ref{fig:fid_mtotal} shows the result.
$\myrec=0.09$ indicates that the GMCMF settle down on the steady state with 
a slight decrement in its total mass
before the calculation ends at $200$ Myr. Contrarily, excessive input $\myrec\gg 0.09$ leads to 
growth of the total mass and do not become steady by $200$Myr as, for example, already seen in Figure~\ref{fig:fid_rec1}.
Less input $\myrec \ll 0.09$ leads to too much gas dispersal into the ISM and eventually 
decreases the total mass in the GMC phase, 
which would not reach any steady state before $200$Myr. 
In Section~\ref{subsec:slope_and_recycle},
we will investigate an analytical justification why $\myrec \sim 0.09$ reaches a steady state
and present the possible range of $\myrec$ that provides a steady GMCMF\@. 

\subsection{The Observed GMCMF Slopes in Arm Regions}
\label{subsec:case8}
\begin{figure}\centering{
\includegraphics[width=0.75\columnwidth,keepaspectratio]{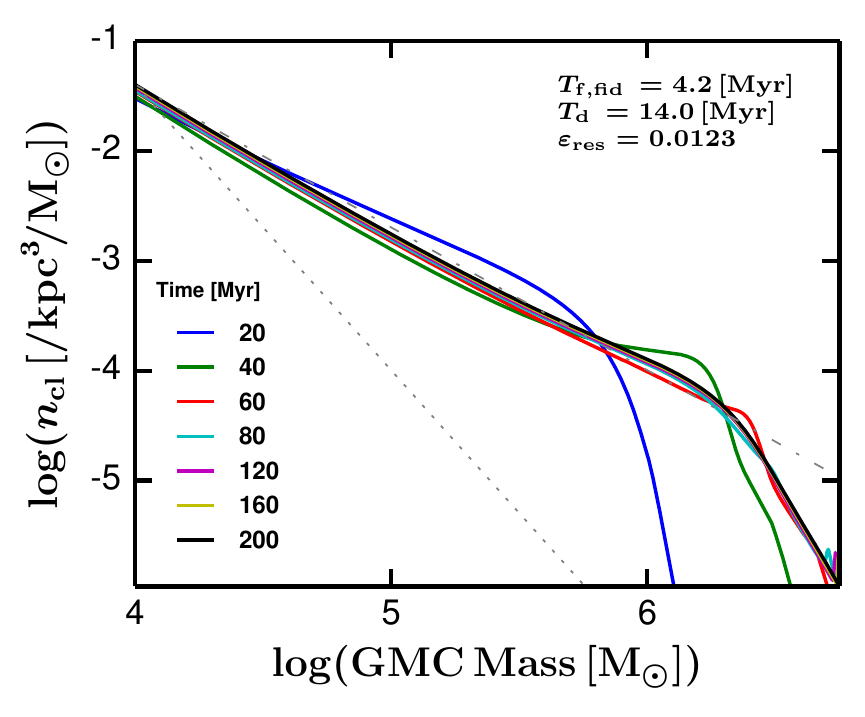}}
\caption{Case 7:
Differential number density $n_{\rm cl}$ as a function of GMC mass
including CCC with $\mytffid=4.2$ Myr and $\myrec=0.0123$.
The color corresponds to the time evolution.
Note that the two colored lines (\ie, the purple line for $120$ Myr 
and the yellow line for $160$ Myr) are overlapped by the black line for $200$ Myr so that
they are not visible on this plot.
As a reference, we plot two gray straight lines; the dot-dashed line is the shallow slope 
($\sim -1.3$) observed in the arm regions, and 
the dotted line represents the steep slope ($\sim -2.6$) observed in the inter-arm regions.
The computed GMCMF successfully reproduces the shallow slope $\sim -1.3$ 
predicted by Equation~(\ref{eq:inutsuka_slope})
and reaches a steady state studied in Figure~\ref{fig:shorttf_mtotal}.}
\label{fig:shorttf_rec5em4}
\end{figure}

\begin{figure}\centering{
\includegraphics[width=0.75\columnwidth,keepaspectratio]{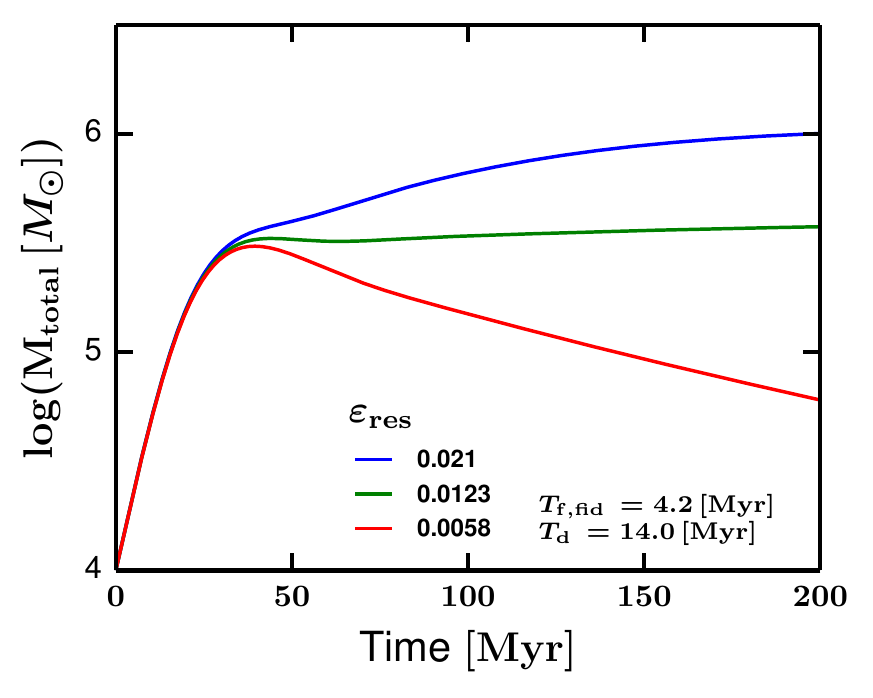}}
\caption{The total mass dependence on $\myrec$ as a function of time with $\mytffid=4.2$ Myr
The different color represents the different $\myrec$ from $5.8 \times 10^{-3}$ to $0.021$. 
The GMCMF with $\myrec=0.0123$ reaches a steady state, $M_{\rm total}$,
which is $\sim 4\times 10^5 \msun$.
$\myrec=0.0123$ reproduces the shallow slope $-\alpha=-1.3$ as shown in Figure~\ref{fig:shorttf_rec5em4}.
Resurrecting more than $\myrec=0.0123$ (\eg, $0.021$ in this figure) increase the total mass in the system. 
Similarly, resurrecting less than $\myrec=0.0123$ (\eg, $5.8 \times 10^{-3}$ in this figure) 
eventually decrease the total mass in the system after 60 Myr. The steady resurrecting factor 
($\sim 1$ per cent) is predictable, which we will explain in Section~\ref{subsec:slope_and_recycle}.}
\label{fig:shorttf_mtotal}
\end{figure}

We now focus on the shorter formation timescale $\mytffid=4.2$ Myr (Case 7)
to examine what $\myrec$ would reproduce the slopes observed in arm regions.
Figure~\ref{fig:shorttf_rec5em4} shows the GMCMF time evolution with $\myrec=0.0123$, which
successfully reproduces the observed shallow slope $\sim-1.3$. Again, to check the steadiness,
we compute the total mass in the system and Figure~\ref{fig:shorttf_mtotal} shows the result
with $\myrec$ from $5.8 \times 10^{-3}$
to 
$0.021$.
The figure also confirms that 
$\myrec=0.0123$
produces a steady state GMCMF\@.
This factor 
$0.0123$
indicates that 
almost
99 per cent of dispersed gas are accreting onto and fueling pre-existing 
GMCs due to the multiple episodes of compression and that only 
1 per cent 
of dispersed gas are turned to form newer generation of GMCs. 
This extreme fraction is expected due to massive GMCs;
massive GMCs have larger surface area than less massive GMCs
and collect large amount of diffuse ISM gas to grow.
Therefore, small $\myrec$, or large $1-\myrec$, is
likely to be realized in arm regions, which have
many massive GMCs.
We will discuss the range of $\myrec$ around $0.0123$ that also keeps the GMCMF steady
in Section~\ref{subsec:slope_and_recycle}.

\subsection{The observed GMCMF Slopes in Inter-Arm Regions}
\label{subsec:case9}
\begin{figure}\centering{
\includegraphics[width=0.75\columnwidth,keepaspectratio]{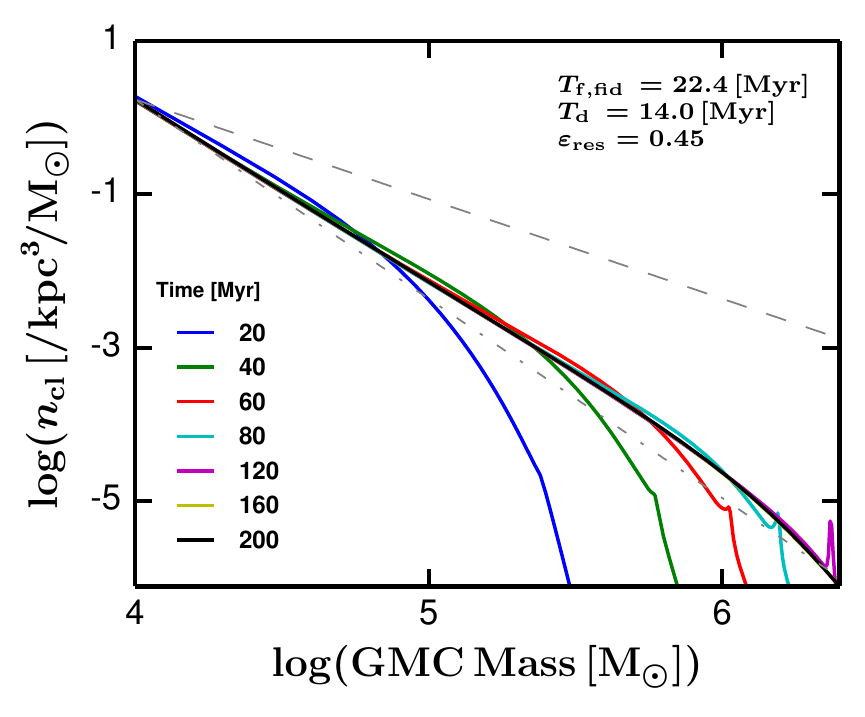}}
\caption{Case 8:
Differential number density $n_{\rm cl}$ as a function of GMC mass
including CCC with $\mytffid=22.4$ Myr and $\myrec=0.45$.
The color corresponds to the time evolution.
Note that the yellow line for $160$ Myr is overlapped by the black line for $200$ Myr,
thus it is not visible in this plot.
As a reference, we plot two gray straight lines; the dot-dashed line is the steep slope 
($\sim -2.6$) observed in the inter-arm regions, and 
the dashed line represents the shallow slope ($\sim -1.3$) observed in the arm regions.
Although the computed GMCMF shows its slope $\sim-2.4$ slightly shallower than the observed value
$-2.6$ due to the coagulation by CCC, this slope is basically predicted by Equation~(\ref{eq:inutsuka_slope})
and the GMCMF reaches a steady state studied in Figure~\ref{fig:longtf_mtotal}.}
\label{fig:longtf_rec038}
\end{figure}

\begin{figure}\centering{
\includegraphics[width=0.75\columnwidth,keepaspectratio]{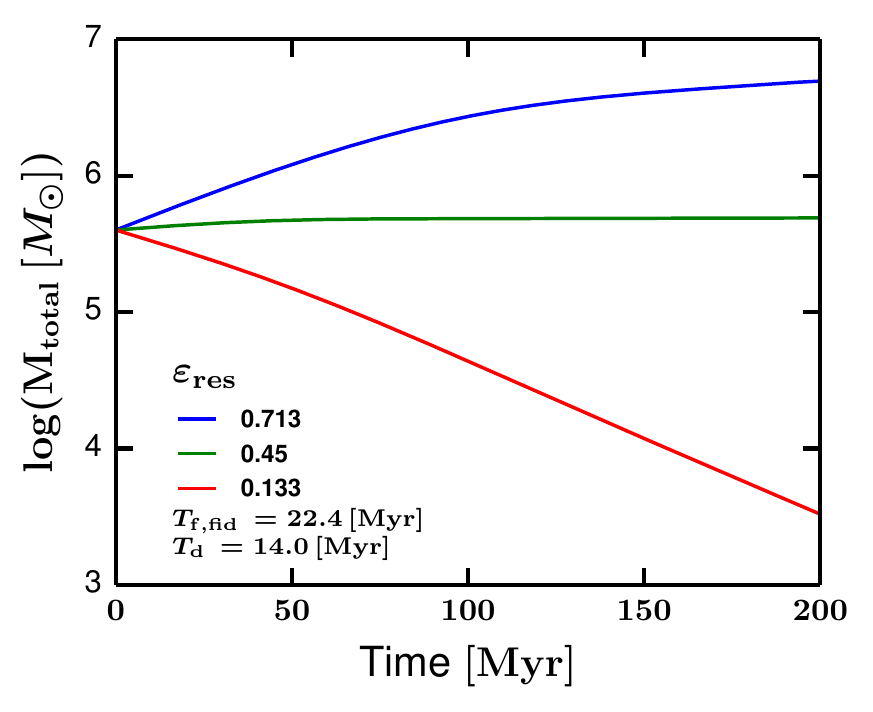}}
\caption{
The total mass dependence on $\myrec$ as a function of time with $\mytffid=22.4$ Myr.
The different color represents the different $\myrec$ from $0.133$ to $0.713$. 
The GMCMF with $\myrec=0.45$ holds 
a steady state $M_{\rm total}$, which is $\sim 5 \times 10^5 \msun$,
and reproduces the steep slope $-\alpha=-2.6$ as shown in Figure~\ref{fig:longtf_rec038}.
The steadiness of the GMCMF with $\myrec=0.45$ indicates that our initial condition 
setting only minimum-mass GMCs is close to the final steady state 
because smaller GMCs dominate the mass budget of the steepened GMCMF\@. 
Resurrecting more than $\myrec=0.45$ (\eg, $0.713$ in this figure) increase the total mass in the system. 
Similarly, resurrecting less than $\myrec=0.45$ (\eg, $0.133$ in this figure) 
decrease the total mass in the system. The steady resurrecting factor 
($\sim 0.45$ per cent) is predictable, which we will explain in Section~\ref{subsec:slope_and_recycle}.}
\label{fig:longtf_mtotal}
\end{figure}

We now focus on the longer formation timescale $\mytffid=22.4$ Myr (Case 8)
to examine what $\myrec$ would reproduce the slopes observed in inter-arm regions.
Figure~\ref{fig:longtf_rec038} shows the GMCMF time evolution with $\myrec=0.45$.
This GMCMF has its slope $\sim -2.4$, shallower than the observed slope $\sim-2.6$
due to the coagulation by CCC, which is 
also observed in Figure~\ref{fig:all_longtf}.
We compute the total mass in the system and Figure~\ref{fig:longtf_mtotal} shows the result
with $\myrec$ from $0.133$ to $0.713$. The figure confirms that 
$\myrec=0.45$ produces a steady state GMCMF\@.
The initial condition is coincidently close enough to the 
steady state with $\myrec=0.45$ thus the GMCMF with $\myrec=0.45$ keep its total mass
over the whole 200 Myr. This factor $0.45$ indicates that 
about 55 per cent of dispersed gas are accreting onto and fueling pre-existing 
GMCs due to the multiple episodes of compression and that 45 per cent 
of dispersed gas are turned to form newer generation of GMCs.
This is naturally expected because the inter-arm regions do not contain many massive GMCs
that may collect diffuse ISM to grow.
We will discuss the range of $\myrec$ around $0.45$ that also keeps the GMCMF steady
in Section~\ref{subsec:slope_and_recycle}.

\subsection{Analytical estimation on the resurrecting factors}
\label{subsec:slope_and_recycle}

In this section, we derive an analytical estimation
for the steady state $\myrec$ with its possible variation as a function of the GMCMF slope
to explain the reason why the values we choose for the resurrecting factor 
in Sections~\ref{subsec:case8} and~\ref{subsec:case9} can successfully reproduce the observed variation of the GMCMFs.
Here, we assume that the GMCMF in a mass range from $\mymmin$ to 
$\mymmax$ has a steady power-law state $n_{\rm cl} = A m^{-\alpha}$,
where $A$ is a constant.
We choose $\mymmin = 10^4 \msun$ 
in our calculation (see Appendix~\ref{subsec:minimum_gmc} for the choice
of $\mymmin$ value), whereas $\mymmax \approx \mtrunc$ as shown in
the figures from our calculation by now.

To derive $\myrec$ in a steady state GMCMF, we multiply $m$ 
by Equation~(\ref{eq:coageq}) without the CCC term 
and modify the dispersal term into a form of $\partial/\partial m$:
\begin{equation}
    \frac{\partial mn_{\rm cl}}{\partial t} 
    + \frac{\partial}{\partial m} F(m) =0 \,.
    \label{eq:coagmasseq}
\end{equation}
Here,
\begin{equation}
    F(m) = \frac{m^2 n_{\rm cl}}{\mytf(m)} - 
    \int_{\mymmin}^{m} \frac{mn_{\rm cl}{\rm d}m}{\mytf(m)} + 
    \int_{\mymmin}^{m} \frac{mn_{\rm cl}{\rm d}m}{\mytd}  \,, 
\end{equation}
is the net mass flux at $m$ across the mass coordinate. 
Because we consider a steady state GMCMF at $m\leq\mtrunc$,
$\mytf$ will be treated as a constant in the following analysis.
Then $F(m)$ can be rewritten as:
\begin{equation}
\begin{aligned}
    F(m) &= A m^{2-\alpha} \left( \frac{1}{T_{\rm f}} -
			\frac{1}{(2-\alpha)T_{\rm f}} +
			\frac{1}{(2-\alpha)T_{\rm d}}
		    \right) \\
         &+ A m_{\rm min}^{2-\alpha} \left(\frac{1}{(2-\alpha)T_{\rm f}} -
            \frac{1}{(2-\alpha)T_{\rm d}}\right) \,. 
\end{aligned}
\end{equation}
This mass flux is useful to evaluate the mass evolution
in the system as we prove from now,
and is also complementary with the original 
number density analysis because $\partial F(m) / \partial m = 0$ 
reproduces the steady state slope
predicted in Equation~(\ref{eq:inutsuka_slope}):
\begin{equation}
    \alpha = 1 + \frac{\mytf}{\mytd} \,.
\end{equation}

We introduce four characteristic
quantities that control $\myrec$; 
the incoming flux at the minimum-mass end, $\myfin$, 
the outgoing flux at the high-mass end, $\myfout$, 
the total mass growth per unit time by all the pre-existing GMCs,
$\mydotmg$, and the total dispersed mass generated per unit time from the system,
$\mydotmd$. They are given as:
\begin{eqnarray}
    \myfin &=& F(\mymmin) = \mymmin^2 n_{\rm cl}/\mytf \,,\label{eq:condition_steady} \\
    \myfout &=& \mymmax^2 n_{\rm cl} / \mytf \,, \\
    \mydotmg &=& \int_{\mymmin}^{\mymmax} \left(n_{\rm cl} m/\mytf\right) {\rm d}m \,, \\
    \mydotmd &=& \int_{\mymmin}^{\mymmax} \left(n_{\rm cl} m/\mytd\right) {\rm d}m \,.
\end{eqnarray}
With these quantities, $F(\mymmax)$ corresponds to $\myfout - \mydotmg + \mydotmd$.
In the steady state, $F(m)$ is independent of $m$. 
Thus $F(m_{\rm min}) = F(m) = F(m_{\rm max})$
and this can be rewritten as:
\begin{equation}
    \myfin = \myfout - \mydotmg + \mydotmd \,.
    \label{eq:equiv_steady} 
\end{equation}

For massive GMCs, the power-law relation $n = A m^{-\alpha}$ is broken:
the treatment is valid if we take $\mymmax \approx \mtrunc$. For $m \ga
\mymmax$, $T_{\rm f}$ is rapidly increases with mass and the dispersal
becomes more important than the growth so that GMCs with $m \ga \mymmax$ are
eventually dispersed. Therefore, $\myfout$ is
interpreted as the mass dispersing rate at $m \geq \mymmax$ and
the total dispersal rate of GMCs is given by $\mydotmd+\myfout$. On the
other hand, $\myfin$ means the formation rate of minimum-mass GMCs. 
Therefore, the resurrecting factor is given by 
\begin{equation}
    \myrec = \frac{\myfin}{\myfout+\mydotmd} \,.
    \label{eq:recdef}
\end{equation}
We consider that the GMCMF from $\mymmin$ to $\mymmax$ becomes in a
steady state when $\myrec = \myrecsteady$.
Combined with Equation~(\ref{eq:equiv_steady}), 
$\myrecsteady$ is given as:
\begin{equation}
\begin{aligned}
    &\myrecsteady \\
    &= \frac{\myfin}{\myfin+\mydotmg} \\
    &=
    \begin{cases}
        &\left\{ 1 + \frac{1}{2-\alpha}
        \left[\left(\frac{\mymmax}{\mymmin}\right)^{2-\alpha} -1\right]
        \right\}^{-1}  \quad (\text{for $\alpha \neq 2$}) \,, \\
        &\left[1+\ln(\mymmax/\mymmin)\right]^{-1} \quad (\text{for $\alpha = 2$}) \,. 
    \end{cases}
    \label{eq:recdef_steady}
\end{aligned}
\end{equation}
Table~\ref{table:epsrec} shows the $\myrecsteady$ computed by
Equation~(\ref{eq:recdef_steady}). Here, as a representative case of $\mymmax<\mtrunc$,
we opt to use $\mymmax=10^{6.2} \msun$, which is also consistent with
the truncated mass scale observed in our computed GMCMF
(see also Fig.~\ref{fig:rec_range}). 
Because $\mymmax \gg \mymmin$, Equation~(\ref{eq:recdef_steady}) can be
simplified as 
\begin{equation}
    \myrecsteady = \\
    \begin{cases}
        &(2-\alpha)\left(\frac{\mymmin}{\mymmax}\right)^{2-\alpha} 
        \quad (\text{for $\alpha<2$}) \,, \\
        &\left[\ln(\mymmax/\mymmin)\right]^{-1} \quad (\text{for $\alpha = 2$}) \,, \\
        &\frac{\alpha-2}{\alpha-1} \quad (\text{for $\alpha>2$}) \,.
    \end{cases}
    \label{eq:simple_res}
\end{equation}
Note that the 
accuracy of the approximation $\mymmax<\mtrunc$ differs between cases,
and thus the computed GMCMFs exhibit the resurrecting factors that slightly deviate
by a factor $0.6-1.2$ from Equation~(\ref{eq:recdef_steady})
(\ie, Table~\ref{table:epsrec}).

\begin{table}
    \caption{Estimated resurrecting factor as a function of growth timescale}
    \centering{
        \begin{tabular}{ccccc}
            \hline
            \hline
            \input{epsrec.table}
            \hline
            \hline
        \end{tabular}
    }\par
    \bigskip
    \textbf{Note.} Predicted $\myrecsteady$ for three timescales: $\mytffid=4.2, 10$, and $22.4$ Myr 
    based on Equation~(\ref{eq:recdef_steady}). All the cases have a constant $\mytd=14$ Myr.
    \label{table:epsrec}
\end{table}

\subsubsection{The range of the resurrecting factor}
\label{subsubsec:myrec_min_max}
The integration of Equation~(\ref{eq:coagmasseq}) over $m$ from $\mymmin$ to
$\mymmax$ results in 
\begin{eqnarray}
    \frac{d M_{\rm total}}{dt} &=& F(\mymmin) - F(\mymmax) \,, \nonumber \\
    &=& \myfin - \myfout - \mydotmd + \mydotmg \,,  \nonumber \\
    &=& \left(1-\frac{1}{\myrec}\right) \myfin + \mydotmg \,.
    \label{eq:mdot_ori} 
\end{eqnarray}
where $M_{\rm total}$ is the total mass of GMCs from $\mymmin$ to $\mymmax$:
\begin{equation}
    M_{\rm total} = \int_{\mymmin}^{\mymmax} m n_{\rm cl} {\rm d}m \,,
\end{equation}
In the steady state, $dM_{\rm total}/dt =0$ and $\myrec =\myrecsteady$. Therefore,
\begin{equation}
    \mydotmg = - \left(1-\frac{1}{\myrecsteady}\right) \myfin \,.
    \label{eq:dmgrow_steady} 
\end{equation}
The variation of $\myrec$ changes $n_{\rm cl}$ and hence $\mydotmg$. 
However for simplicity, we here ignore the dependence of $\mydotmg$ on $\myrec$
by considering $\myrec \sim \myrecsteady$.
From Equations~(\ref{eq:mdot_ori}) and~(\ref{eq:dmgrow_steady}), we obtain
\begin{equation}
   \frac{d M_{\rm total}}{dt} = \left(\frac{1}{\myrecsteady}
   -\frac{1}{\myrec}\right) F_{\rm in} \,.
\end{equation}

If the GMCMF is steady and the total mass in the system $M_{\rm total}$ 
does not increase/reduce by factor $\beta$ within a certain timescale $\mytsteady$,
\begin{equation}
    \left| \frac{\beta M_{\rm total}}{\dot M_{\rm total}} \right| \gtrsim \mytsteady \,.
\end{equation}
Combining with Equation~(\ref{eq:dmgrow_steady}), 
the range of $\myrec$ that leads to steady state becomes:
\begin{equation}
   \left( 1- \frac{\myrecsteady}{\myrecsteady+a} \right) \lesssim \frac{\myrec}{\myrecsteady}
   \lesssim \left( 1+ \frac{\myrecsteady}{a-\myrecsteady} \right) \,,
   \label{eq:rec_range}
\end{equation}
where $a=\myfin \mytsteady/(M_{\rm total} \beta)$.

\begin{figure}\centering{
\includegraphics[width=0.75\columnwidth,keepaspectratio]{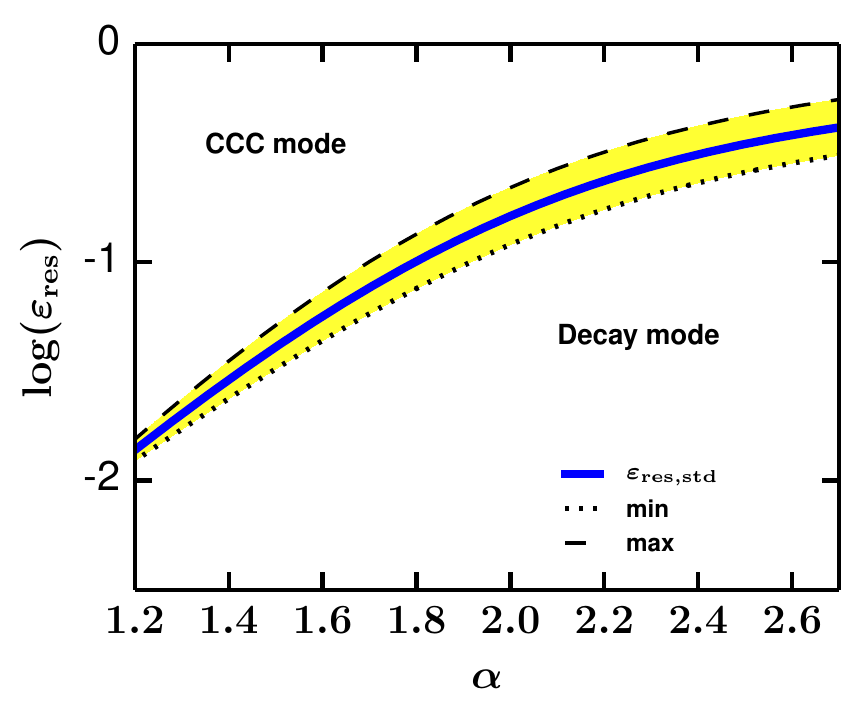}}
\caption{The range of the resurrecting factor, $\myrec$, that sustain a steady GMCMF
as a function of the GMCMF slope $-\alpha$ (\ie, $n_{\rm cl}(m) \propto m^{-\alpha}$).
The thick solid blue line shows the estimated $\myrecsteady$ based on Equation~(\ref{eq:recdef_steady}).
The dotted and dashed lines correspond to minimum and maximum of $\myrec$ evaluated from
Equation~(\ref{eq:rec_range}) with $\beta=0.5$, $\mytsteady=100$ Myr, and $\mymmax=10^{6.2}\msun$.
$\myrec$ increases with $\alpha$ and saturates when $\alpha$ becomes large (\ie, when the GMCMF has a steep slope)
as expected from Equation~(\ref{eq:simple_res}).
The variation of $\myrec$ at a given $\alpha$ is less than one order of magnitude 
for all the $\alpha$ range.
The GMCMFs with shallow slopes typically have $\myrec = \mathcal{O}(10^{-2})$ whereas
$\mathcal{O}(10^{-1})$ for steep slopes, which indicates that dispersed gas contributes
more to the formation of newer generation GMC in inter-arm regions.
The overloading resurrection more than maximum $\myrec$ leads to the CCC-dominated regime.
On the other hand, the GMCMF decays with the resurrection less than the minimum $\myrec$.}
\label{fig:rec_range}
\end{figure}

Figure~\ref{fig:rec_range} shows the $\myrecsteady$ and $\myrec$ variation based on 
Equations~(\ref{eq:recdef_steady}) and~(\ref{eq:rec_range}) with $\mymmax=10^{6.2}\msun$. 
Here, we assume that 
the GMCMF can be steady and localized within the half galactic rotation (in a case of 
two spiral arms in a galactic disk) so that $\mytsteady = 100$ Myr and opt to
use $\beta=0.5$. The figure shows that $\myrec$ increase with $\alpha$.
This trend is naturally expected because the inter-arm regions have less number of 
massive GMCs that can sweep up dispersed gas compared with the arm regions
and dispersed gas easily produce minimum-mass GMCs when they experience 
multiple episodes of compression.

On the one hand, overwhelming resurrection more than the maximum $\myrec$ increase the 
number density of GMCs, which ends up with the CCC-dominated regime after the long time elapse.
On the other hand, the GMC number density decreases so that the GMCMF decays
with resurrecting factor less than the minimum $\myrec$.

\subsubsection{Unseen gas}
\label{subsubsec:analytic_summary}
Up until now, we have shown that $\myrec$ between $\mathcal{O}(0.01)$ to $\mathcal{O}(0.1)$ 
reproduces the observed GMCMF slope. The estimated face values themselves are important,
but moreover, our results strongly suggest that understanding the fate of dispersed gas is inevitable
to study the gas resurrecting processes in the ISM\@. The gas phases that are not well observed yet 
(\eg, CO-dark H$_2$ gas \citep{Hosokawa2006b,Tang2016,Xu2016} and optically thick H{\sc i} gas \citep{Fukui2015a}) 
also come into play as well as usual CO-bright molecular gas. 
Three dimensional detailed magnetohydrodynamics simulation, for example, is required to understand the evolution of
those unseen gas phases and to test whether it reproduces $\myrec$ that we predict here.

\section{DISCUSSION}
\label{subsec:discussion}
\subsection{CCC frequency and its impact}
\label{subsec:dichotomy}
Our results indicate that CCC is limited only at the massive-end and does not alter the GMCMF 
time evolution significantly. This seems contrary to the CCC importance 
suggested by galactic simulations and observations, but is still consistent.
The collision timescale typically varies from 1 to 100 Myr (see Figure~\ref{fig:tcol_all_support}),
which is consistent with global galactic simulations (\cf, 25 Myr by \citet[]{Tasker2009}, 
8 - 28 Myr by \citet[]{Dobbs2015}).
Figure~\ref{fig:tcol_all_support} shows that GMCs $>10^6 \msun$ experience CCC 
much more frequently than GMCs $<10^5 \msun$ as expected by the kernel function
(Equation~(\ref{eq:kernel})). The lack of star cluster formation during CCC
in our calculation generates many massive GMCs at the massive-end \citep[\cf,][]{Tasker2009}.
However, we expect that such drastic star cluster formation take place
only when larger GMCs collide each other \citep[]{Fukui2014}, which 
may affect the massive-end but not the slope. 
Therefore, our results indicate that 
inclusion of star cluster formation still does not impact the GMCMF slope significantly.

\subsection{CCC-driven phase}
\label{subsec:eq_or_noneq}
In this section, we are going to investigate whether or not
the cascade collisional coagulation alone can create a steady state GMCMF,
which we investigated so far.
Let us consider the mass flux as a function of GMC mass $m$.
Suppose that the CCC kernel function has its mass dependence as $K\propto m^p$ and 
the resultant differential number density has a spectra $n\propto m^{-\alpha}$,
the mass flux $F$ due to the cascade collisional coagulation becomes
\citep[\cf,][]{Kobayashi2010}:
\begin{equation}
    F \propto m^3 n^2 K  \propto m^{-2\alpha + p + 3} \,.
    \label{eq:massflux_col}
\end{equation}
In a steady state, $F$ is constant across the whole mass range (\ie $F\propto m^0$),
therefore 
\begin{equation}
    p = 2\alpha-3 \,.
\end{equation} 
If the mass dependence of the kernel function has a variation with $p \in [-0.4, 2.2]$
the observed slopes can be reproduced, which typically varies $-\alpha = -1.3 - -2.6$.
Thus, the regions where the GMC number density is large (\eg, the Galactic Center)
may 
form this type of CCC-driven GMCMF \citep[\cf,][]{Tsuboi2015}.
However, the collision alone reduce the number of minimum-mass GMCs quickly
and also create infinitely massive GMCs. Therefore,
we still need some proper prescription for the creation process at the low-mass end
and the dispersal process at the high-mass end. We will investigate 
such a formulation in our upcoming paper.
Also note that, the GMCMF may have the same slope 
when collision is not cascade but always occurs only with the minimum-mass bin,
which we will also report in our forthcoming paper.

\section{CONCLUSIONS AND SUMMARY}
\label{sec:summary}
We have calculated the GMCMF time evolution by formulating a coagulation equation
to reproduce and explain the possible origin of the observed variation in the GMCMF slopes.
Our formulation is based on the paradigm that GMCs are created from magnetized WNM
in galactic disks through multiple episodes of compression 
and such compression is driven by the network of expanding H{\sc ii} regions
and the late phase of supernova remnants.

Contrary to the previous works, we revealed that GMC self-growth 
overwhelm CCC\@. The GMCMF slope varies according to
the diversity of GMC self-growth/dispersal timescales on galactic scales,
which is essentially pointed out by \citet[]{Inutsuka2015};
shallow slope with smaller ratio of growth/dispersal timescale.
Our results demonstrate that future large radio surveys
are capable of putting constraints on GMC self-growth
and dispersal timescales in different environment on galactic scales.

We further include the gas resurrection process to regenerate
GMCs after GMCs become dispersed due to star formation.
The mass flux analysis under a steady state GMCMF assumption provides
the analytical relation between the gas resurrecting factor 
and the GMC self-growth/dispersal timescales.
Both the calculated GMCMF and the analytical relation agrees that
typically over 90 \% of the dispersed gas accrete onto the pre-existing GMCs
in arm regions whereas 
only half in inter-arm regions.
Typically, the arm regions have shorter formation timescale ($\sim 4$ Myr)
and the inter-arm regions have longer formation timescale ($\sim 22$ Myr),
indicating that the rate of massive star formation and/or supernovae 
is 
higher in arm regions than inter-arm regions.

The observational statistics for low mass GMCs are currently limited by completeness.
Our results suggest that the GMCMF has a single power-law exponent 
in the mass range $<10^{5.5} \msun$, which may be revealed 
once observations start to resolve those smaller GMCs.

Due to its short evolution timescale ($\lesssim 100$ Myr), 
our formulation can be widely applied to the GMCMFs 
across galactic disks.
Although galactic morphological diversity (\eg elliptical and dwarf) is not explicitly
described and currently beyond our scope,
observed steep slope in elliptical galaxies \citep[\eg][]{Utomo2015} indicates 
the potential capability of our formulation for various galactic morphologies.
The CCC-dominated phase still remains to be studied and we 
are planning to report another formulation in our forthcoming paper
that investigates the GMCMF in higher number density regions (such as the Galactic Center).
Lastly, star formation phase of ISM is not studied yet in detail in the current article. 
Our formulation can be essentially extended to compute 
massive star and star cluster mass function and examine the Schmidt conjecture
(\ie, Kennicutt-Schmidt law), which we would like to report as well in our 
forthcoming paper.

\section*{ACKNOWLEDGMENTS}
We are grateful to the anonymous referee for providing thoughtful comments, 
which improved our manuscript in great details.
MINK (15J04974), SI (23244027, 16H02160), 
and HK (26287101) are supported by Grants-in-Aid from the Ministry of Education, Culture,
Sports, Science, and Technology of Japan. 
KH is supported by a grant from National Observatory of Japan.
This work was supported by the Astrobiology Center Project of the National Institute of 
Natural Sciences (NINS) (grant Number AB271020).
MINK thank Yasuo Fukui, Akihiko Hirota, Hidetoshi Sano, and Yusuke Hattori for educating us
with observational backgrounds. MINK also appreciate Tsuyoshi Inoue, Hosokawa Takashi, Masahiro Nagashima,
Martin Bureau, Ralph Schoenrich, Steven Longmore, Toby Moore, Jonathan Henshaw, 
Gary Fuller, Anthony Whitworth, Nicolas Peretto, Ana Duarte Cabral, Paul Clark, 
Cathie Clarke, Thomas Haworth, Chiaki Kobayashi, and Hua-Bai Li for fruitful discussion.
\par


\begin{appendices}

\section{VARIOUS EFFECTS}
\label{sec:variouseffects}
In the following appendices, we examine a variety of effects that modify the time evolution of the GMCMF;
(i) Variation in the cross section, (ii) Variation in the relative velocities,
(iii) Initial conditions, (iv) The choice for the minimum-mass GMC.
The first two effects ((i) and (ii)) increase and decrease the rate for collisional coagulation.
The setup variations fit into (iii) and (iv).

\section{Variation in the collision rate: cross section (geometrical structure and gravitational focusing) 
and relative velocity}
\label{subsec:var_col}
In Section~\ref{subsec:ccc_term}, we introduced the CCC kernel function, 
which has a correction factor $c_{\rm col}$. This $c_{\rm col}$ involves
various effects; three main effects can be the geometrical structure, the gravitational focusing, 
and the relative velocity variation. We will explore these effects to verify
that $c_{\rm col}$ has an order unity and $c_{\rm col}=1$ is enough approximation.

\subsection{Geometrical structure}
\label{subsubsec:geometrics}
For simplicity, let us ignore the thickness of GMCs, which makes $c_{\rm col}$ a factor larger,
and assume that GMCs always coagulate even when their peripheries alone touch each other. 

\begin{figure}\centering{
\includegraphics[width=0.75\columnwidth,keepaspectratio]{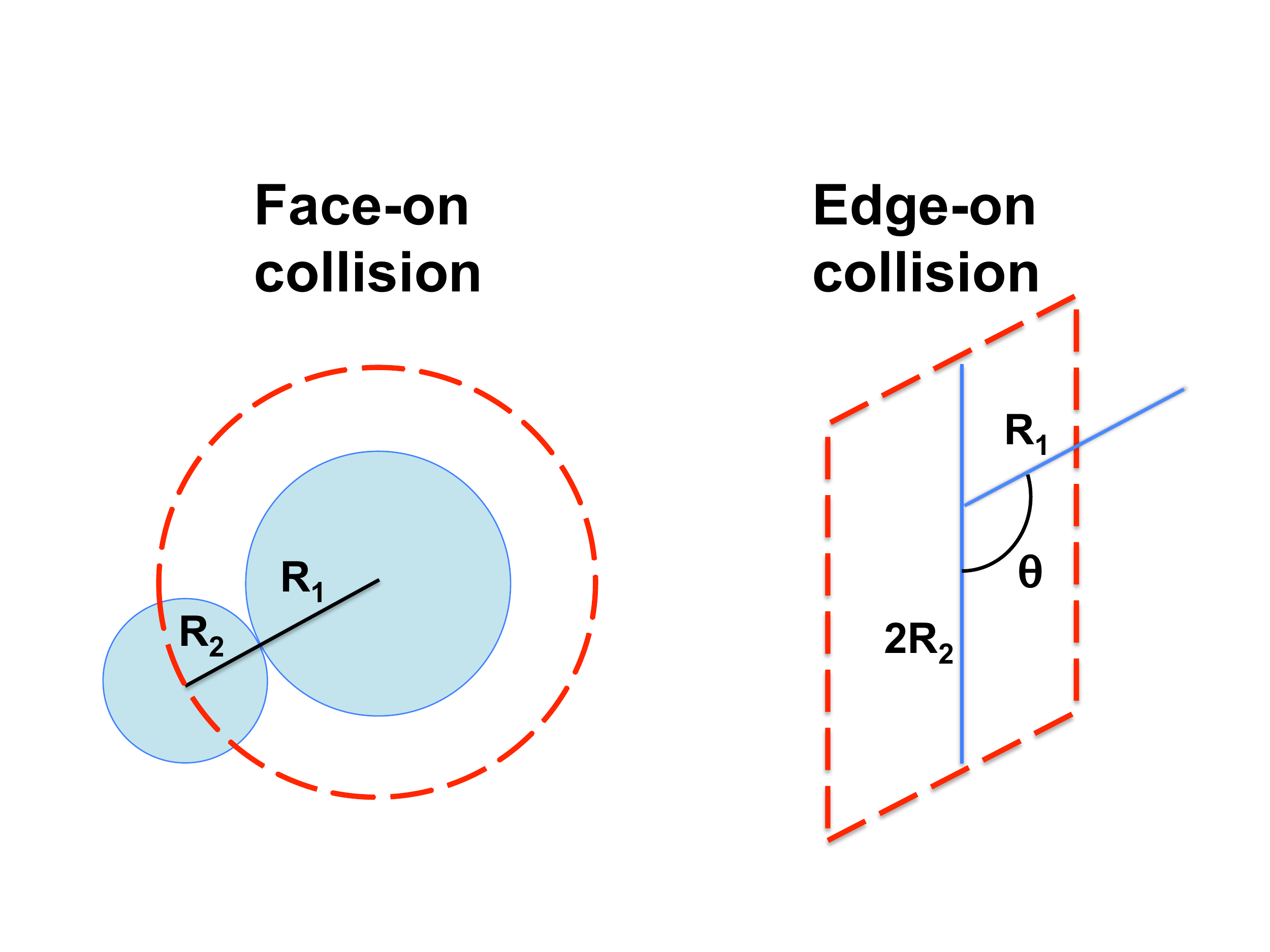}}
\caption{Schematic figure describing the geometrical configuration during cloud-cloud collisions
(left: face-on collision; right: edge-on collision). GMCs have mass $m_1$ and $m_2$, radii $R_1$ and $R_2$
respectively. The red dashed area corresponds to the collisional cross section $\sigma_{\rm col \, peri \, 1,2}$.}
\label{fig:sch_col}
\end{figure}

Given that GMCs have a uniform column density $\Sigma_{\rm mol}$, 
the cross section for face-on collisional coagulation between GMCs whose masses are $m_1$ and $m_2$ 
($m_1 \geq m_2$), $\sigma_{\rm col \, peri \, 1,2}$, is described as:
\begin{equation}
\begin{aligned}
    \sigma_{\rm col \, peri \, 1,2} &= \pi \left( R_1 + R_2 \right)^2 \\
    &= \pi \left( \sqrt{\frac{m_1}{\pi \Sigma_{\rm mol}}} + \sqrt{\frac{m_2}{\pi \Sigma_{\rm mol}}}\right)^2 \\
    &= \frac{\left( \sqrt{m_1} + \sqrt{m_2} \right)^2}{\Sigma_{\rm mol}} \,,
\end{aligned}
\end{equation}
where $R_1$ and $R_2$ are the radii of GMCs with masses $m_1$ and $m_2$ respectively
(see the left panel of Figure~\ref{fig:sch_col} for the schematic configuration).
In this case, the correction factor becomes:
\begin{equation}
    c_{\rm col} = \frac{\sigma_{\rm col \, peri \, 1,2}}{\sigma_{\rm col \, 1,2}} 
    = 1+ \frac{2\sqrt{m_1}\sqrt{m_2}}{m_1 + m_2} \,.
    \label{eq:peri}
\end{equation}
The second term gives the deviation from the unity and thus $c_{\rm col}$ has its maximum value $2$ when $m_1 = m_2$.
In reality, GMCs are less likely to coagulate together when their peripheries alone touch each other
so that Equation~(\ref{eq:peri}) overestimates the collision frequency 
and $c_{\rm col}$ should be smaller by some factor.
Therefore, we can assume $c_{\rm col} = 1$ for simplicity.

The edge-on collision can be similarly evaluated. Supposed that GMCs collide with an angle $\theta$,
the cross section becomes $2R_1 \times 2R_2 \times \sin \theta$ (see the right panel of 
Figure~\ref{fig:sch_col}). $\sigma_{\rm col \, peri \, 1,2}$ can be obtained by 
averaging this cross section over $\theta$:
\begin{equation}
\begin{aligned}
    \sigma_{\rm col \, peri \, 1,2} &= \frac{\int_{0}^{\pi} 2R_1 2R_2 \sin \theta}{\pi} \\
    &= \frac{8}{\pi}R_1 R_2 \\
    &= \frac{8}{\pi^2 \Sigma_{\rm mol}} \sqrt{m_1 m_2}\,.
\end{aligned}
\end{equation}
Therefore, the correction factor becomes:
\begin{equation}
    c_{\rm col} = \frac{8}{\pi^2} \frac{\sqrt{m_1 m_2}}{m_1 + m_2} \,,
\end{equation}
whose maximum is $0.57$ when $m_1 = m_2$. Therefore, edge-on collision 
is less frequent than our fiducial cross section in Equation~(\ref{eq:kernel}).
Less frequent collision may impact on the shape of the massive end of the GMCMF\@.
However, the slope calculated with overestimated $c_{\rm col}=1$ is not affected by
CCC as shown in the figures by now.
Therefore, for simplicity we can assume $c_{\rm col}=1$ even for the edge-on collision
if we restrict ourselves to the slope of the GMCMF.
\par

One caveat here is that we focus only on the collision in face-on and edge-on.
The other orientations leads to the cross section smaller than the configuration we analyze here 
and would not greatly modify our GMCMF calculation results. Thus, we ignore such orientations
for simplicity and left them for other works. 
Further investigation, especially three-dimensional hydrodynamics simulation, is needed.

\subsection{Gravitational focusing factor}
\label{subsubsec:gravfactor}
During a close-by encounter, two GMCs experience gravitational attraction toward each other 
(\ie gravitational focusing effect), which effectively enlarges the cross section 
so that the rate for collisional coagulation increases. To evaluate the gravitational focusing
effect, let us consider an example where a GMC whose mass and radius are $m_2$ and $r_2$
initially flies at a speed $v_{\rm i}$ with an impact parameter $b$ with respect to 
another GMC with mass $m_1$ and radius $r_1$. 
Combining the angular momentum conservation and energy conservation, one can derive 
the ratio of the effective cross section $b^2$ against the area $(r_1 + r_2)^2$ as:
\begin{equation}
\begin{aligned}
    \frac{b^2}{\left( r_1 + r_2 \right)^2} &= 1 + \Theta \\
                                           &= 1 + \frac{v_{\rm esc}^2}{v_{\rm i}^2} \\
                                           &= 1 + \frac{2{\rm G}(m_1+m_2)}{(r_1+r_2)v_{\rm i}^2} \,,
\end{aligned}
\end{equation}
where $v_{\rm esc}$ is the escape velocity when the GMC peripheries touch each other.
$\Theta$ is so called gravitational focusing factor, which characterize the relative increment of 
cross section due to gravitational attraction.
\par
The condition where the gravitational focusing effect outnumbers the cross section would be
$\Theta \gtrsim 1$. This condition with our ordinal assumption that GMCs have constant column density $\Sigma$ and 
$v_{\rm i} = 10 \, \mathrm{km \, s^{-1}}$ leads to:
\begin{equation}
\begin{aligned}
    &\frac{m_1 + m_2}{r_1 + r_2} \gtrsim 1.16 \times 10^4 \, \mathrm{\msun \, pc^{-1}} \\
    &\Rightarrow \frac{m_1 + m_2}{\sqrt{m_1} + \sqrt{m_2}} \gtrsim 
    \frac{1.16 \times 10^4 \, \mathrm{\msun \, pc^{-1}} }{\sqrt{\pi \Sigma}} \,.
    \label{eq:gravf}
\end{aligned}
\end{equation}
The left hand side of Equation~(\ref{eq:gravf}) has its minimum when $m_2 = 0.17 m_1$ and maximum
when $m_2 = m_1$. The most conservative estimation can be obtained by employing the maximum case $m_2 = m_1$,
which results in $m_1 \gtrsim 5 \times 10^6 \msun$. 
This lower boundary may increase by another factor 3 to 5 because other collisions with $m_2 < m_1$ generally exist
and also $v_{\rm esc} \gtrsim \mathcal{O}(10) v_{\rm i}$ is required for CCC to 
overwhelm GMC self-growth/dispersal. Therefore, we can conclude that the gravitational focusing effect
may be involved only in the mass range beyond $10^7 \msun$. 
As shown in Sections~\ref{sec:results1} and after that, all the GMCMFs have already exhibited 
their well-defined slope below $10^7 \msun$ thus treating the correction factor $c_{\rm col} = 1$ is 
an enough approximation to compute the GMCMF slope correctly.
\par
Note that our fiducial cross section is given as the sum of geometrical cross sections (\ie $\pi r_1^2 + \pi r_2^2$) 
rather than $\pi (r_1 + r_2)^2$ that we studied here. However, the former is smaller than the latter
by only a factor one to two, thus the criterion $m_1 \gtrsim 5 \times 10^6 \msun$ remains valid in our setup.
Overall, the gravitational focusing effect can increase the cross section by some factors but we can still
assume $c_{\rm col}=1$ for the mass range defining the GMCMF slope.

\subsection{Relative velocity}
\label{subsubsec:relvel}
The relative velocity between clouds is another component that may vary the collisional kernel.
Because our model assumes that GMCs are swept up by expanding shells (\eg H{\sc ii} regions and supernova remnants),
we expect that the relative velocity between GMCs is comparable to a typical shock expanding speed 
driven by the ionization-dissociation front \citep[]{Hosokawa2006b} and does not have 
any strong mass dependence. Even a simple setup can derive this estimation; 
the Rankine-Hugoniot relation \citep[see][]{Landau1959}, for example,
predicts that the velocity change across the shock front is described as
$v_2 = v_1 \times \{(\gamma -1)\mathcal{M}_1^2 +2\} \times ((\gamma +1)\mathcal{M}_1^2)^{-1}$,
where $v_1$ and $v_2$ are the fluid speed in the pre-shock and post-shock regions respectively,
$\mathcal{M}_1$ is the Mach number in the pre-shock region, and $\gamma$ is the polytropic index
for the fluid. With a strong shock $\mathcal{M}\gg1$, the cooling time can be longer than 
the dynamical time for the shock passing through a GMC, which leads to $\gamma>1$ rather than
$\gamma=1$ even for a molecular cloud. Then the velocity change becomes $v_2 = \mathcal{O}(0.1) \, v_1$ 
thus GMCs in the post-shock regions may have a relative velocity $\mathcal{O}(0.1) \times 10 \, \mathrm{km \, s^{-1}}$.
Semi-analytic studies also demonstrate that 
that a molecular cloud moves slowly than the shock itself \citep[]{Iwasaki2011a, Iwasaki2011b}, 
thus the GMC relative velocity would be somewhat smaller than $10 \, \mathrm{km \, s^{-1}}$. 
Indeed, observations show that the mean intercloud dispersion in the Galaxy
is a few $\mathrm{km \, s^{-1}}$ \citep[][]{stark1989,stark2005,stark2006}.
Therefore, we opt to use $V_{\rm rel,0} = 10 \, \mathrm{km \, s^{-1}}$ irrespective of GMC mass,
which may reduce $c_{\rm col}$ by a factor of a few. This compensates the increasing trend
discussed in the previous sections so that $c_{\rm col}$ remains the order of unity.

Note that the shear motion in galactic disk would not outnumber the 
velocity dispersion produced by shocks. For example, 
supposed that two clouds at $R$ in a galactocentric coordinates
have a separation of $d$ and the speed of the galactic rotation 
is $V_{\rm rot}$, then the shear speed can be estimated about
$\sim V_{\rm rot}  d  R^{-1}$. With $V_{\rm rot} = 200 \, \mathrm{km \, s^{-1}}$,
$d = 100 \, {\rm pc}$, and $R = 8 \mathrm{kpc}$ (\eg, solar neighbors),
the shear speed becomes $2.5 \, \mathrm{km \, s^{-1}}$.
At the galactic centers with smaller $R$, the shear motion definitely contributes more 
to the GMC relative velocity, but the significance is non-trivial
because GMCs are closer each other than disk regions so that may have smaller $d$.
Because we now focus on disk regions, the shear motion 
does not alter the relative velocity appreciably
in the conditions we present in the current article.

\section{Initial conditions}
\label{subsec:init}
\begin{figure}\centering{
\includegraphics[width=0.75\columnwidth,keepaspectratio]{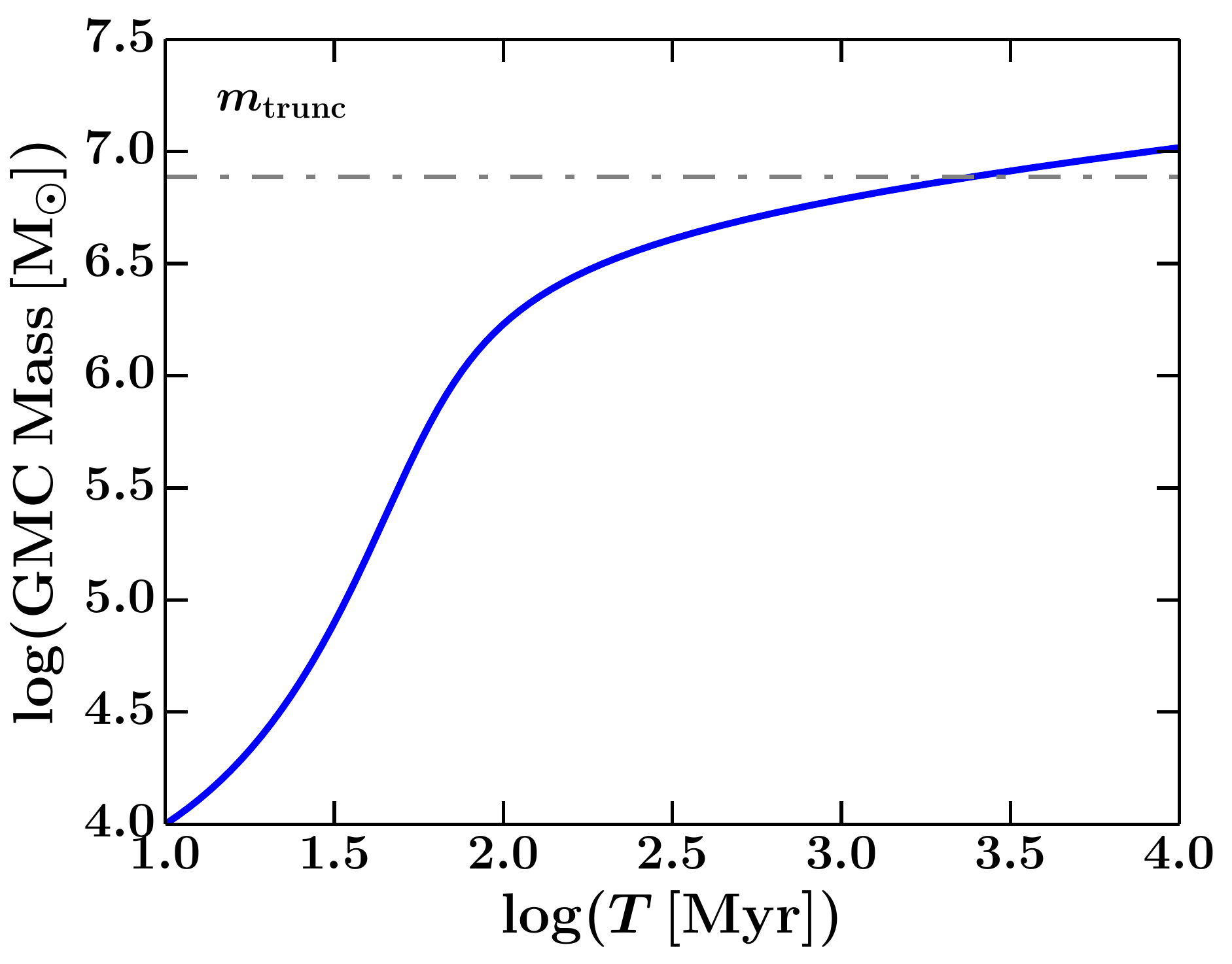}}
\caption{$T(m) = \int \mytf(m)/m \ {\rm d}m$ with $\mytffid=10$ Myr, which gives 
the typical self-growth timescales for GMCs that have not undergone CCC\@. 
A horizontal dash-dot line indicates $\mtrunc$.
As expected from the exponential growth picture in Section~\ref{subsec:growth},
$T(m)$ logarithmically grows until GMCs acquire $\mtrunc/\gamma \sim 10^6 \msun$ 
up to which $\mytf$ is almost constant (\ie, $\mytffid$).
$T(m)$ rapidly increase beyond $\mtrunc/\gamma$ due to the $\mytf$ dependence on
$m$ (see Equation~(\ref{eq:myratio_mdep})). }
\label{fig:tgrow}
\end{figure}

In general, the GMCMF time evolution may depend on the initial condition. 
To test the importance of the initial condition, we compute the time
that is required for minimum-mass GMCs to grow to mass $m$ by H{\sc i} cloud 
accretion as:
\begin{equation}
    T(m) = \int_{\mymmin}^{m} \left[\mytf (m)/m \right]\, {\rm d}m \,.
    \label{eq:time_m}
\end{equation}
This integration originates from our substitution $({\rm d}m/{\rm d}t)_{\rm self} = m/\mytf (m)$
in Section~\ref{subsec:growth}. Equation~(\ref{eq:time_m}) suggests that GMCs grow exponentially 
in $m<\mtrunc$. We take the integration in Equation~(\ref{eq:time_m}) and Figure~\ref{fig:tgrow} shows the result 
as a function of GMC mass. Note that, here, we include the $\mytf$ dependence on GMC mass 
(see Equation~(\ref{eq:myratio_mdep})). Therefore, the computed time exhibits a rapid increase
at the massive end in Figure~\ref{fig:tgrow}.

As shown in Figure~\ref{fig:tgrow}, minimum-mass GMCs reach $10^6 \msun$
within $100$ Myrs. Therefore, initial conditions for our calculation is erased typically less than one 
rotation of galactic disk and we expect that the observed GMCMF would not be affected significantly
by the initial conditions. In case the initial condition is not influential in the resultant GMCMF,
then formulation that depends on the initial condition 
\citep[\eg, cosmological Press-Schechter formalism:][]{Press1974}
may not describe well the time evolution of the GMCMF.

Note that the GMCMF evolution can be modified in earlier stage if with the initial condition 
where very massive GMCs dominate the total mass of the system. 
However, based on the multiphase magnetohydrodynamics simulations \citep[\eg][]{Inoue2008}, 
multiple successive compression should first create smaller GMCs and cultivate them to massive ones,
rather than produce very massive ones directly from WNM\@. Therefore, we opt to preset only minimum-mass GMCs 
at the beginning of our calculation (\cf Appendix~\ref{subsec:minimum_gmc} for how to determine
the minimum mass).

\section{The choice for the minimum-mass GMC}
\label{subsec:minimum_gmc}
The concept of minimum-mass ($\mymmin$) GMC is necessary to compute the GMCMF numerically.
However, it is not trivial what value should be used as the ``minimum'' mass.
Thus, to estimate this minimum value, let us consider a large hydrogen cloud.
Given the mean number density in this cloud is $\sim 10^2 \, \mathrm{cm^{-3}}$ and 
the visual extinction is 1 or larger due to the dust shielding to form molecular hydrogen
(\ie the total hydrogen column density $2 \times 10^{21} \, \mathrm{cm^{-2}}$ 
\citep{Bohlin1978,Rachford2009,Draine2011text}),
the typical length for this type of hydrogen cloud is expected about $2\times10^{21}/10^2 = 2\times10^{19} \mathrm{cm}$.
If we assume that the cloud has this length stretched at all the solid angles, 
then the total hydrogen mass in this cloud is given as:
\begin{equation}
\begin{aligned}
    &\frac{4}{3} \pi \left( 2\times 10^{19} \right)^3 \times 10^2 \times 2\mu m_{\rm H}^{} \\
    &\simeq 2.6 \times 10^4 \msun \,.
    \label{eq:mmin}
\end{aligned}
\end{equation}
ISM simulations \citep[\eg,][]{Inoue2012} also indicate that $10^4 \msun$ is a typical
minimum GMC\@ created after 10 Myr. 
The minimum limit $\gtrsim10^4 \msun$ is consistent with the value 
observationally used \citep[see][]{Williams2000} as the definition of GMCs. 
Thus we opt to use $10^4 \msun$ as the minimum-mass for GMCs.
\par
However, the estimation by Equation~(\ref{eq:mmin}) evaluates the minimum amount of total hydrogen 
but not molecular hydrogen alone.
Therefore, in a static case where, for example, diffuse H{\sc i} gas extend around dense H$_2$ gas, 
the mass of dense H$_2$ gas can be half or less than Equation~(\ref{eq:mmin}) prediction.
Under a dynamical situation, turbulence stirs H{\sc i} and H$_2$ gas to break up H$_2$ cloud into 
smaller H$_2$ clouds whose mass can be smaller than that estimated by Equation~(\ref{eq:mmin}).
Moreover, if molecular gas self-shielding alone
is effective when the dust is poor,
the required total hydrogen column density can be smaller by a few orders of magnitude.
Thus, even though we cannot specify the exact number,
it maybe
optimal to set somewhat smaller value than shown in Equation~(\ref{eq:mmin}) as the minimum-mass of GMCs.
We will investigate the minimum-mass choice further 
in our forthcoming paper, combining with the effect CO-dark H$_2$ gas and the
$\mytd$ variation in low-mass less star-forming GMCs.
\par
Note that, if the highest-density (star forming) H$_2$ regions have a filament structure
\citep[\cf][]{Andre2010,Molinari2010,Arzoumanian2011,Hill2011,Konyves2015},
it is not trivial whether H{\sc i} gas sufficiently extend around those H$_2$ regions
to sustain the column density condition ($2 \times 10^{21} \, \mathrm{cm^{-2}}$).
This needs to be studied further and left for future works.

\section{The mass flow timescale}
\label{sec:tflow}
$\mytf$ defined in Equation~(\ref{eq:coageq}) are the timescales for 
the {\it number} flow. However $\mytcol$ defined in Equation~(\ref{eq:tcol}) is 
the timescale for the {\it mass} flow. To compare these two timescales,
we need to convert $\mytf$ to the mass flow timescale, which can be evaluated as:
\begin{equation}
\begin{aligned}
    T_{\rm f, \, mass \, flow} &= \left. mn_{\rm cl} \middle/ \frac{\partial}{\partial m} \left( \frac{m^2 n_{\rm cl}}{\mytf(m)} - 
    \int_{\mymmin}^{m}\frac{mn_{\rm cl} {\rm d}m}{\mytf(m)}\right)  \right. \\
    &= \left. mn_{\rm cl} \middle/ \left( \frac{\partial}{\partial m} \left( \frac{m^2 n_{\rm cl}}{\mytf(m)} \right) - 
    \frac{mn_{\rm cl} {\rm d}m}{\mytf(m)} \right) \right. \,,
\end{aligned}
\end{equation}
(see Equation~(\ref{eq:coagmasseq})). 
The gray line in Figure~(\ref{fig:tcol_all_support}) shows this $T_{\rm f, \, mass \, flow}$
instead of $\mytf$.

\end{appendices}


\bibliographystyle{apj}
\bibliography{galaxy}



\end{document}

%% file: cases.table
Cases & \multicolumn{5}{c}{Parameters}\\
& $\mytffid$  &  $\mytd$  &  Cloud-Cloud Collisions  &  $\myrec$  &  Figures\\
& $($ Myr $)$ &  $($ Myr $)$ & & & \\
 \hline 
 \hline
1   & 10               & 14  & yes     & support          &  Fig.~\ref{fig:all_support} \\ 
2   & 10               & 14  & no      & support          &  Fig.~\ref{fig:self_support} \\ 
3   & 4.2              & 14  & yes     & support          &  Fig.~\ref{fig:all_shorttf} \\ 
4   & 22.4             & 14  & yes     & support          &  Fig.~\ref{fig:all_longtf} \\ 
5   & 10               & 14  & yes     & 1                &  Fig.~\ref{fig:fid_rec1} \\ 
6   & 10               & 14  & yes     & 0.09             &  Fig.~\ref{fig:fid_rec001} \\ 
7   & 4.2              & 14  & yes     & 0.0123           &  Fig.~\ref{fig:shorttf_rec5em4} \\ 
8   & 22.4             & 14  & yes     & 0.45             &  Fig.~\ref{fig:longtf_rec038} \\ 

%% file: epsrec.table
                & \multicolumn{3}{c}{Estimated $\myrec$}\\
$\log(\mymmax)$  & $\mytffid=4.2$  & $\mytffid=10$  & $\mytffid=22.4$ \\  
  $($ \msun $)$  & $($ Myr $)$ & $($ Myr $)$ & $($ Myr $)$ \\
 \hline 
 \hline
6.5 & 0.0201 & 0.0798 & 0.383 \\

%% file: mink2017_gmcmf_arxiv.bbl
\begin{thebibliography}{}
\expandafter\ifx\csname natexlab\endcsname\relax\def\natexlab#1{#1}\fi

\bibitem[{{Andr{\'e}} {et~al.}(2011){Andr{\'e}}, {Men'shchikov}, {K{\"o}nyves},
  \& {Arzoumanian}}]{Andre2011}
{Andr{\'e}}, P., {Men'shchikov}, A., {K{\"o}nyves}, V., \& {Arzoumanian}, D.
  2011, in IAU Symposium, Vol. 270, Computational Star Formation, ed.
  J.~{Alves}, B.~G. {Elmegreen}, J.~M. {Girart}, \& V.~{Trimble}, 255--262

\bibitem[{{Andr{\'e}} {et~al.}(2010){Andr{\'e}}, {Men'shchikov}, {Bontemps},
  {K{\"o}nyves}, {Motte}, {Schneider}, {Didelon}, {Minier}, {Saraceno},
  {Ward-Thompson}, {di Francesco}, {White}, {Molinari}, {Testi}, {Abergel},
  {Griffin}, {Henning}, {Royer}, {Mer{\'{\i}}n}, {Vavrek}, {Attard},
  {Arzoumanian}, {Wilson}, {Ade}, {Aussel}, {Baluteau}, {Benedettini},
  {Bernard}, {Blommaert}, {Cambr{\'e}sy}, {Cox}, {di Giorgio}, {Hargrave},
  {Hennemann}, {Huang}, {Kirk}, {Krause}, {Launhardt}, {Leeks}, {Le Pennec},
  {Li}, {Martin}, {Maury}, {Olofsson}, {Omont}, {Peretto}, {Pezzuto}, {Prusti},
  {Roussel}, {Russeil}, {Sauvage}, {Sibthorpe}, {Sicilia-Aguilar}, {Spinoglio},
  {Waelkens}, {Woodcraft}, \& {Zavagno}}]{Andre2010}
{Andr{\'e}}, P., {Men'shchikov}, A., {Bontemps}, S., {et~al.} 2010, \aap, 518,
  L102

\bibitem[{{Arzoumanian} {et~al.}(2011){Arzoumanian}, {Andr{\'e}}, {Didelon},
  {K{\"o}nyves}, {Schneider}, {Men'shchikov}, {Sousbie}, {Zavagno}, {Bontemps},
  {di Francesco}, {Griffin}, {Hennemann}, {Hill}, {Kirk}, {Martin}, {Minier},
  {Molinari}, {Motte}, {Peretto}, {Pezzuto}, {Spinoglio}, {Ward-Thompson},
  {White}, \& {Wilson}}]{Arzoumanian2011}
{Arzoumanian}, D., {Andr{\'e}}, P., {Didelon}, P., {et~al.} 2011, \aap, 529, L6

\bibitem[{{Audit} \& {Hennebelle}(2005)}]{Audit2005}
{Audit}, E., \& {Hennebelle}, P. 2005, \aap, 433, 1

\bibitem[{{Audit} \& {Hennebelle}(2008)}]{Audit2008}
{Audit}, E., \& {Hennebelle}, P. 2008, in Astronomical Society of the Pacific
  Conference Series, Vol. 385, Numerical Modeling of Space Plasma Flows, ed.
  N.~V. {Pogorelov}, E.~{Audit}, \& G.~P. {Zank}, 73

\bibitem[{{Bohlin} {et~al.}(1978){Bohlin}, {Savage}, \& {Drake}}]{Bohlin1978}
{Bohlin}, R.~C., {Savage}, B.~D., \& {Drake}, J.~F. 1978, \apj, 224, 132

\bibitem[{{Colombo} {et~al.}(2014{\natexlab{a}}){Colombo}, {Hughes},
  {Schinnerer}, {Meidt}, {Leroy}, {Pety}, {Dobbs}, {Garc{\'{\i}}a-Burillo},
  {Dumas}, {Thompson}, {Schuster}, \& {Kramer}}]{Colombo2014a}
{Colombo}, D., {Hughes}, A., {Schinnerer}, E., {et~al.} 2014{\natexlab{a}},
  \apj, 784, 3

\bibitem[{{Colombo} {et~al.}(2014{\natexlab{b}}){Colombo}, {Meidt},
  {Schinnerer}, {Garc{\'{\i}}a-Burillo}, {Hughes}, {Pety}, {Leroy}, {Dobbs},
  {Dumas}, {Thompson}, {Schuster}, \& {Kramer}}]{Colombo2014b}
{Colombo}, D., {Meidt}, S.~E., {Schinnerer}, E., {et~al.} 2014{\natexlab{b}},
  \apj, 784, 4

\bibitem[{{Corder} {et~al.}(2008){Corder}, {Sheth}, {Scoville}, {Koda},
  {Vogel}, \& {Ostriker}}]{Corder2008}
{Corder}, S., {Sheth}, K., {Scoville}, N.~Z., {et~al.} 2008, \apj, 689, 148

\bibitem[{{Dobbs} \& {Pringle}(2013)}]{Dobbs2013}
{Dobbs}, C.~L., \& {Pringle}, J.~E. 2013, \mnras, 432, 653

\bibitem[{{Dobbs} {et~al.}(2015){Dobbs}, {Pringle}, \&
  {Duarte-Cabral}}]{Dobbs2015}
{Dobbs}, C.~L., {Pringle}, J.~E., \& {Duarte-Cabral}, A. 2015, \mnras, 446,
  3608

\bibitem[{{Draine}(2011)}]{Draine2011text}
{Draine}, B.~T. 2011, {Physics of the Interstellar and Intergalactic Medium}

\bibitem[{{Engargiola} {et~al.}(2003){Engargiola}, {Plambeck}, {Rosolowsky}, \&
  {Blitz}}]{Engargiola2003}
{Engargiola}, G., {Plambeck}, R.~L., {Rosolowsky}, E., \& {Blitz}, L. 2003,
  \apjs, 149, 343

\bibitem[{{Field} {et~al.}(1969){Field}, {Goldsmith}, \& {Habing}}]{Field1969}
{Field}, G.~B., {Goldsmith}, D.~W., \& {Habing}, H.~J. 1969, \apjl, 155, L149

\bibitem[{{Fukui} {et~al.}(2001){Fukui}, {Mizuno}, {Yamaguchi}, {Mizuno}, \&
  {Onishi}}]{Fukui2001}
{Fukui}, Y., {Mizuno}, N., {Yamaguchi}, R., {Mizuno}, A., \& {Onishi}, T. 2001,
  \pasj, 53, L41

\bibitem[{{Fukui} {et~al.}(2015{\natexlab{a}}){Fukui}, {Torii}, {Onishi},
  {Yamamoto}, {Okamoto}, {Hayakawa}, {Tachihara}, \& {Sano}}]{Fukui2015a}
{Fukui}, Y., {Torii}, K., {Onishi}, T., {et~al.} 2015{\natexlab{a}}, \apj, 798,
  6

\bibitem[{{Fukui} {et~al.}(2014){Fukui}, {Ohama}, {Hanaoka}, {Furukawa},
  {Torii}, {Dawson}, {Mizuno}, {Hasegawa}, {Fukuda}, {Soga}, {Moribe},
  {Kuroda}, {Hayakawa}, {Kawamura}, {Kuwahara}, {Yamamoto}, {Okuda}, {Onishi},
  {Maezawa}, \& {Mizuno}}]{Fukui2014}
{Fukui}, Y., {Ohama}, A., {Hanaoka}, N., {et~al.} 2014, \apj, 780, 36

\bibitem[{{Fukui} {et~al.}(2015{\natexlab{b}}){Fukui}, {Harada}, {Tokuda},
  {Morioka}, {Onishi}, {Torii}, {Ohama}, {Hattori}, {Nayak}, {Meixner},
  {Sewi{\l}o}, {Indebetouw}, {Kawamura}, {Saigo}, {Yamamoto}, {Tachihara},
  {Minamidani}, {Inoue}, {Madden}, {Galametz}, {Lebouteiller}, {Mizuno}, \&
  {Chen}}]{Fukui2015b}
{Fukui}, Y., {Harada}, R., {Tokuda}, K., {et~al.} 2015{\natexlab{b}}, \apjl,
  807, L4

\bibitem[{{Fukui} {et~al.}(2016){Fukui}, {Torii}, {Ohama}, {Hasegawa},
  {Hattori}, {Sano}, {Ohashi}, {Fujii}, {Kuwahara}, {Mizuno}, {Dawson},
  {Yamamoto}, {Tachihara}, {Okuda}, {Onishi}, \& {Mizuno}}]{Fukui2016}
{Fukui}, Y., {Torii}, K., {Ohama}, A., {et~al.} 2016, \apj, 820, 26

\bibitem[{{Gratier} {et~al.}(2012){Gratier}, {Braine}, {Rodriguez-Fernandez},
  {Schuster}, {Kramer}, {Corbelli}, {Combes}, {Brouillet}, {van der Werf}, \&
  {R{\"o}llig}}]{Gratier2012}
{Gratier}, P., {Braine}, J., {Rodriguez-Fernandez}, N.~J., {et~al.} 2012, \aap,
  542, A108

\bibitem[{{Heitsch} {et~al.}(2005){Heitsch}, {Burkert}, {Hartmann}, {Slyz}, \&
  {Devriendt}}]{Heitsch2005}
{Heitsch}, F., {Burkert}, A., {Hartmann}, L.~W., {Slyz}, A.~D., \& {Devriendt},
  J.~E.~G. 2005, \apjl, 633, L113

\bibitem[{{Heitsch} {et~al.}(2006){Heitsch}, {Slyz}, {Devriendt}, {Hartmann},
  \& {Burkert}}]{Heitsch2006b}
{Heitsch}, F., {Slyz}, A.~D., {Devriendt}, J.~E.~G., {Hartmann}, L.~W., \&
  {Burkert}, A. 2006, \apj, 648, 1052

\bibitem[{{Heitsch} {et~al.}(2009){Heitsch}, {Stone}, \&
  {Hartmann}}]{Heitsch2009}
{Heitsch}, F., {Stone}, J.~M., \& {Hartmann}, L.~W. 2009, \apj, 695, 248

\bibitem[{{Hennebelle} \& {Audit}(2007)}]{Hennebelle2007a}
{Hennebelle}, P., \& {Audit}, E. 2007, \aap, 465, 431

\bibitem[{{Hennebelle} {et~al.}(2007){Hennebelle}, {Audit}, \&
  {Miville-Desch{\^e}nes}}]{Hennebelle2007b}
{Hennebelle}, P., {Audit}, E., \& {Miville-Desch{\^e}nes}, M.-A. 2007, \aap,
  465, 445

\bibitem[{{Hennebelle} \& {P{\'e}rault}(1999)}]{Hennebelle1999}
{Hennebelle}, P., \& {P{\'e}rault}, M. 1999, \aap, 351, 309

\bibitem[{{Hennebelle} \& {P{\'e}rault}(2000)}]{Hennebelle2000}
---. 2000, \aap, 359, 1124

\bibitem[{{Heyer} \& {Dame}(2015)}]{Heyer2015}
{Heyer}, M., \& {Dame}, T.~M. 2015, \araa, 53, 583

\bibitem[{{Heyer} {et~al.}(2001){Heyer}, {Carpenter}, \& {Snell}}]{Heyer2001}
{Heyer}, M.~H., {Carpenter}, J.~M., \& {Snell}, R.~L. 2001, \apj, 551, 852

\bibitem[{{Hill} {et~al.}(2011){Hill}, {Motte}, {Didelon}, {Bontemps},
  {Minier}, {Hennemann}, {Schneider}, {Andr{\'e}}, {Men'shchikov}, {Anderson},
  {Arzoumanian}, {Bernard}, {di Francesco}, {Elia}, {Giannini}, {Griffin},
  {K{\"o}nyves}, {Kirk}, {Marston}, {Martin}, {Molinari}, {Nguyen Luong},
  {Peretto}, {Pezzuto}, {Roussel}, {Sauvage}, {Sousbie}, {Testi},
  {Ward-Thompson}, {White}, {Wilson}, \& {Zavagno}}]{Hill2011}
{Hill}, T., {Motte}, F., {Didelon}, P., {et~al.} 2011, \aap, 533, A94

\bibitem[{{Hosokawa} \& {Inutsuka}(2006)}]{Hosokawa2006b}
{Hosokawa}, T., \& {Inutsuka}, S.-i. 2006, \apj, 646, 240

\bibitem[{{Inoue} \& {Inutsuka}(2008)}]{Inoue2008}
{Inoue}, T., \& {Inutsuka}, S.-i. 2008, \apj, 687, 303

\bibitem[{{Inoue} \& {Inutsuka}(2009)}]{Inoue2009}
---. 2009, \apj, 704, 161

\bibitem[{{Inoue} \& {Inutsuka}(2012)}]{Inoue2012}
---. 2012, \apj, 759, 35

\bibitem[{{Inutsuka}(2001)}]{Inutsuka2001}
{Inutsuka}, S.-i. 2001, \apjl, 559, L149

\bibitem[{{Inutsuka} {et~al.}(2015){Inutsuka}, {Inoue}, {Iwasaki}, \&
  {Hosokawa}}]{Inutsuka2015}
{Inutsuka}, S.-i., {Inoue}, T., {Iwasaki}, K., \& {Hosokawa}, T. 2015, \aap,
  580, A49

\bibitem[{{Iwasaki} {et~al.}(2011{\natexlab{a}}){Iwasaki}, {Inutsuka}, \&
  {Tsuribe}}]{Iwasaki2011a}
{Iwasaki}, K., {Inutsuka}, S.-i., \& {Tsuribe}, T. 2011{\natexlab{a}}, \apj,
  733, 16

\bibitem[{{Iwasaki} {et~al.}(2011{\natexlab{b}}){Iwasaki}, {Inutsuka}, \&
  {Tsuribe}}]{Iwasaki2011b}
---. 2011{\natexlab{b}}, \apj, 733, 17

\bibitem[{{Kawamura} {et~al.}(2009){Kawamura}, {Mizuno}, {Minamidani},
  {Filipovi{\'c}}, {Staveley-Smith}, {Kim}, {Mizuno}, {Onishi}, {Mizuno}, \&
  {Fukui}}]{Kawamura2009}
{Kawamura}, A., {Mizuno}, Y., {Minamidani}, T., {et~al.} 2009, \apjs, 184, 1

\bibitem[{{Kennicutt} \& {Evans}(2012)}]{Kennicutt2012}
{Kennicutt}, R.~C., \& {Evans}, N.~J. 2012, \araa, 50, 531

\bibitem[{{Kobayashi} \& {Tanaka}(2010)}]{Kobayashi2010}
{Kobayashi}, H., \& {Tanaka}, H. 2010, \icarus, 206, 735

\bibitem[{{Koda} {et~al.}(2009){Koda}, {Scoville}, {Sawada}, {La Vigne},
  {Vogel}, {Potts}, {Carpenter}, {Corder}, {Wright}, {White}, {Zauderer},
  {Patience}, {Sargent}, {Bock}, {Hawkins}, {Hodges}, {Kemball}, {Lamb},
  {Plambeck}, {Pound}, {Scott}, {Teuben}, \& {Woody}}]{Koda2009}
{Koda}, J., {Scoville}, N., {Sawada}, T., {et~al.} 2009, \apjl, 700, L132

\bibitem[{{Koda} {et~al.}(2011){Koda}, {Sawada}, {Wright}, {Teuben}, {Corder},
  {Patience}, {Scoville}, {Donovan Meyer}, \& {Egusa}}]{Koda2011}
{Koda}, J., {Sawada}, T., {Wright}, M.~C.~H., {et~al.} 2011, \apjs, 193, 19

\bibitem[{{Koda} {et~al.}(2012){Koda}, {Scoville}, {Hasegawa}, {Calzetti},
  {Donovan Meyer}, {Egusa}, {Kennicutt}, {Kuno}, {Louie}, {Momose}, {Sawada},
  {Sorai}, \& {Umei}}]{Koda2012}
{Koda}, J., {Scoville}, N., {Hasegawa}, T., {et~al.} 2012, \apj, 761, 41

\bibitem[{{K{\"o}nyves} {et~al.}(2015){K{\"o}nyves}, {Andr{\'e}},
  {Men'shchikov}, {Palmeirim}, {Arzoumanian}, {Schneider}, {Roy}, {Didelon},
  {Maury}, {Shimajiri}, {Di Francesco}, {Bontemps}, {Peretto}, {Benedettini},
  {Bernard}, {Elia}, {Griffin}, {Hill}, {Kirk}, {Ladjelate}, {Marsh}, {Martin},
  {Motte}, {Nguy{\^e}n Luong}, {Pezzuto}, {Roussel}, {Rygl}, {Sadavoy},
  {Schisano}, {Spinoglio}, {Ward-Thompson}, \& {White}}]{Konyves2015}
{K{\"o}nyves}, V., {Andr{\'e}}, P., {Men'shchikov}, A., {et~al.} 2015, \aap,
  584, A91

\bibitem[{{K{\"o}rtgen} \& {Banerjee}(2015)}]{Kortgen2015}
{K{\"o}rtgen}, B., \& {Banerjee}, R. 2015, \mnras, 451, 3340

\bibitem[{{Koyama} \& {Inutsuka}(2000)}]{Koyama2000}
{Koyama}, H., \& {Inutsuka}, S.-I. 2000, \apj, 532, 980

\bibitem[{{Koyama} \& {Inutsuka}(2002)}]{Koyama2002}
{Koyama}, H., \& {Inutsuka}, S.-i. 2002, \apjl, 564, L97

\bibitem[{{Kramer} {et~al.}(1998){Kramer}, {Stutzki}, {Rohrig}, \&
  {Corneliussen}}]{Kramer1998}
{Kramer}, C., {Stutzki}, J., {Rohrig}, R., \& {Corneliussen}, U. 1998, \aap,
  329, 249

\bibitem[{{Kwan}(1979)}]{Kwan1979}
{Kwan}, J. 1979, \apj, 229, 567

\bibitem[{{Landau} \& {Lifshitz}(1959)}]{Landau1959}
{Landau}, L.~D., \& {Lifshitz}, E.~M. 1959, {Fluid mechanics}

\bibitem[{{Madau} \& {Dickinson}(2014)}]{Madau2014}
{Madau}, P., \& {Dickinson}, M. 2014, \araa, 52, 415

\bibitem[{{McKee} \& {Ostriker}(1977)}]{McKee1977}
{McKee}, C.~F., \& {Ostriker}, J.~P. 1977, \apj, 218, 148

\bibitem[{{Meidt} {et~al.}(2015){Meidt}, {Hughes}, {Dobbs}, {Pety}, {Thompson},
  {Garc{\'{\i}}a-Burillo}, {Leroy}, {Schinnerer}, {Colombo}, {Querejeta},
  {Kramer}, {Schuster}, \& {Dumas}}]{Meidt2015}
{Meidt}, S.~E., {Hughes}, A., {Dobbs}, C.~L., {et~al.} 2015, \apj, 806, 72

\bibitem[{{Molinari} {et~al.}(2010){Molinari}, {Swinyard}, {Bally}, {Barlow},
  {Bernard}, {Martin}, {Moore}, {Noriega-Crespo}, {Plume}, {Testi}, {Zavagno},
  {Abergel}, {Ali}, {Anderson}, {Andr{\'e}}, {Baluteau}, {Battersby},
  {Beltr{\'a}n}, {Benedettini}, {Billot}, {Blommaert}, {Bontemps}, {Boulanger},
  {Brand}, {Brunt}, {Burton}, {Calzoletti}, {Carey}, {Caselli}, {Cesaroni},
  {Cernicharo}, {Chakrabarti}, {Chrysostomou}, {Cohen}, {Compiegne}, {de
  Bernardis}, {de Gasperis}, {di Giorgio}, {Elia}, {Faustini}, {Flagey},
  {Fukui}, {Fuller}, {Ganga}, {Garcia-Lario}, {Glenn}, {Goldsmith}, {Griffin},
  {Hoare}, {Huang}, {Ikhenaode}, {Joblin}, {Joncas}, {Juvela}, {Kirk},
  {Lagache}, {Li}, {Lim}, {Lord}, {Marengo}, {Marshall}, {Masi}, {Massi},
  {Matsuura}, {Minier}, {Miville-Desch{\^e}nes}, {Montier}, {Morgan}, {Motte},
  {Mottram}, {M{\"u}ller}, {Natoli}, {Neves}, {Olmi}, {Paladini}, {Paradis},
  {Parsons}, {Peretto}, {Pestalozzi}, {Pezzuto}, {Piacentini}, {Piazzo},
  {Polychroni}, {Pomar{\`e}s}, {Popescu}, {Reach}, {Ristorcelli}, {Robitaille},
  {Robitaille}, {Rod{\'o}n}, {Roy}, {Royer}, {Russeil}, {Saraceno}, {Sauvage},
  {Schilke}, {Schisano}, {Schneider}, {Schuller}, {Schulz}, {Sibthorpe},
  {Smith}, {Smith}, {Spinoglio}, {Stamatellos}, {Strafella}, {Stringfellow},
  {Sturm}, {Taylor}, {Thompson}, {Traficante}, {Tuffs}, {Umana}, {Valenziano},
  {Vavrek}, {Veneziani}, {Viti}, {Waelkens}, {Ward-Thompson}, {White},
  {Wilcock}, {Wyrowski}, {Yorke}, \& {Zhang}}]{Molinari2010}
{Molinari}, S., {Swinyard}, B., {Bally}, J., {et~al.} 2010, \aap, 518, L100

\bibitem[{{Nakamura} {et~al.}(2012){Nakamura}, {Miura}, {Kitamura},
  {Shimajiri}, {Kawabe}, {Akashi}, {Ikeda}, {Tsukagoshi}, {Momose}, {Nishi}, \&
  {Li}}]{Nakamura2012}
{Nakamura}, F., {Miura}, T., {Kitamura}, Y., {et~al.} 2012, \apj, 746, 25

\bibitem[{{Onishi} {et~al.}(1999){Onishi}, {Kawamura}, {Abe}, {Yamaguchi},
  {Saito}, {Moriguchi}, {Mizuno}, {Ogawa}, \& {Fukui}}]{Onishi1999}
{Onishi}, T., {Kawamura}, A., {Abe}, R., {et~al.} 1999, \pasj, 51, 871

\bibitem[{{Pety} {et~al.}(2013){Pety}, {Schinnerer}, {Leroy}, {Hughes},
  {Meidt}, {Colombo}, {Dumas}, {Garc{\'{\i}}a-Burillo}, {Schuster}, {Kramer},
  {Dobbs}, \& {Thompson}}]{Pety2013}
{Pety}, J., {Schinnerer}, E., {Leroy}, A.~K., {et~al.} 2013, \apj, 779, 43

\bibitem[{{Press} \& {Schechter}(1974)}]{Press1974}
{Press}, W.~H., \& {Schechter}, P. 1974, \apj, 187, 425

\bibitem[{{Rachford} {et~al.}(2009){Rachford}, {Snow}, {Destree}, {Ross},
  {Ferlet}, {Friedman}, {Gry}, {Jenkins}, {Morton}, {Savage}, {Shull},
  {Sonnentrucker}, {Tumlinson}, {Vidal-Madjar}, {Welty}, \&
  {York}}]{Rachford2009}
{Rachford}, B.~L., {Snow}, T.~P., {Destree}, J.~D., {et~al.} 2009, \apjs, 180,
  125

\bibitem[{{Rice} {et~al.}(2016){Rice}, {Goodman}, {Bergin}, {Beaumont}, \&
  {Dame}}]{Rice2016}
{Rice}, T.~S., {Goodman}, A.~A., {Bergin}, E.~A., {Beaumont}, C., \& {Dame},
  T.~M. 2016, \apj, 822, 52

\bibitem[{{Rosolowsky} {et~al.}(2003){Rosolowsky}, {Engargiola}, {Plambeck}, \&
  {Blitz}}]{Rosolowsky2003}
{Rosolowsky}, E., {Engargiola}, G., {Plambeck}, R., \& {Blitz}, L. 2003, \apj,
  599, 258

\bibitem[{{Rosolowsky} {et~al.}(2007){Rosolowsky}, {Keto}, {Matsushita}, \&
  {Willner}}]{Rosolowsky2007}
{Rosolowsky}, E., {Keto}, E., {Matsushita}, S., \& {Willner}, S.~P. 2007, \apj,
  661, 830

\bibitem[{{Roy} {et~al.}(2015){Roy}, {Andr{\'e}}, {Arzoumanian}, {Peretto},
  {Palmeirim}, {K{\"o}nyves}, {Schneider}, {Benedettini}, {Di Francesco},
  {Elia}, {Hill}, {Ladjelate}, {Louvet}, {Motte}, {Pezzuto}, {Schisano},
  {Shimajiri}, {Spinoglio}, {Ward-Thompson}, \& {White}}]{Roy2015}
{Roy}, A., {Andr{\'e}}, P., {Arzoumanian}, D., {et~al.} 2015, \aap, 584, A111

\bibitem[{{Salpeter}(1955)}]{Salpeter1955}
{Salpeter}, E.~E. 1955, \apj, 121, 161

\bibitem[{{Schinnerer} {et~al.}(2013){Schinnerer}, {Meidt}, {Pety}, {Hughes},
  {Colombo}, {Garc{\'{\i}}a-Burillo}, {Schuster}, {Dumas}, {Dobbs}, {Leroy},
  {Kramer}, {Thompson}, \& {Regan}}]{Schinnerer2013}
{Schinnerer}, E., {Meidt}, S.~E., {Pety}, J., {et~al.} 2013, \apj, 779, 42

\bibitem[{{Schinnerer} {et~al.}(2017){Schinnerer}, {Meidt}, {Colombo},
  {Chandar}, {Dobbs}, {Garcia-Burillo}, {Hughes}, {Leroy}, {Pety}, {Querejeta},
  {Kramer}, \& {Schuster}}]{Schinnerer2017}
{Schinnerer}, E., {Meidt}, S.~E., {Colombo}, D., {et~al.} 2017, ArXiv e-prints
  (accepted to \apj), arXiv:1701.02184

\bibitem[{{Schruba} {et~al.}(2011){Schruba}, {Leroy}, {Walter}, {Bigiel},
  {Brinks}, {de Blok}, {Dumas}, {Kramer}, {Rosolowsky}, {Sandstrom},
  {Schuster}, {Usero}, {Weiss}, \& {Wiesemeyer}}]{Schruba2011}
{Schruba}, A., {Leroy}, A.~K., {Walter}, F., {et~al.} 2011, \aj, 142, 37

\bibitem[{{Scoville} \& {Hersh}(1979)}]{Scoville1979}
{Scoville}, N.~Z., \& {Hersh}, K. 1979, \apj, 229, 578

\bibitem[{{Stark} \& {Brand}(1989)}]{stark1989}
{Stark}, A.~A., \& {Brand}, J. 1989, \apj, 339, 763

\bibitem[{{Stark} \& {Lee}(2005)}]{stark2005}
{Stark}, A.~A., \& {Lee}, Y. 2005, \apjl, 619, L159

\bibitem[{{Stark} \& {Lee}(2006)}]{stark2006}
---. 2006, \apjl, 641, L113

\bibitem[{{Tacconi} {et~al.}(2010){Tacconi}, {Genzel}, {Neri}, {Cox}, {Cooper},
  {Shapiro}, {Bolatto}, {Bouch{\'e}}, {Bournaud}, {Burkert}, {Combes},
  {Comerford}, {Davis}, {Schreiber}, {Garcia-Burillo}, {Gracia-Carpio}, {Lutz},
  {Naab}, {Omont}, {Shapley}, {Sternberg}, \& {Weiner}}]{Tacconi2010}
{Tacconi}, L.~J., {Genzel}, R., {Neri}, R., {et~al.} 2010, \nat, 463, 781

\bibitem[{{Tachihara} {et~al.}(2000){Tachihara}, {Mizuno}, \&
  {Fukui}}]{Tachihara2000}
{Tachihara}, K., {Mizuno}, A., \& {Fukui}, Y. 2000, \apj, 528, 817

\bibitem[{{Tan}(2000)}]{Tan2000}
{Tan}, J.~C. 2000, \apj, 536, 173

\bibitem[{{Tang} {et~al.}(2016){Tang}, {Li}, {Heiles}, {Wang}, {Pan}, \&
  {Wang}}]{Tang2016}
{Tang}, N., {Li}, D., {Heiles}, C., {et~al.} 2016, \aap, 593, A42

\bibitem[{{Tasker} \& {Tan}(2009)}]{Tasker2009}
{Tasker}, E.~J., \& {Tan}, J.~C. 2009, \apj, 700, 358

\bibitem[{{Tomisaka}(1986)}]{Tomisaka1986}
{Tomisaka}, K. 1986, \pasj, 38, 95

\bibitem[{{Torii} {et~al.}(2015){Torii}, {Hasegawa}, {Hattori}, {Sano},
  {Ohama}, {Yamamoto}, {Tachihara}, {Soga}, {Shimizu}, {Okuda}, {Mizuno},
  {Onishi}, {Mizuno}, \& {Fukui}}]{Torii2015}
{Torii}, K., {Hasegawa}, K., {Hattori}, Y., {et~al.} 2015, \apj, 806, 7

\bibitem[{{Tosaki} {et~al.}(2016){Tosaki}, {Kohno}, {Harada}, {Tanaka},
  {Egusa}, {Izumi}, {Takano}, {Nakajima}, {Taniguchi}, \&
  {Tamura}}]{Tosaki2016}
{Tosaki}, T., {Kohno}, K., {Harada}, N., {et~al.} 2016, ArXiv e-prints (in
  press at \pasj), arXiv:1612.00948

\bibitem[{{Tsuboi} {et~al.}(2015){Tsuboi}, {Miyazaki}, \&
  {Uehara}}]{Tsuboi2015}
{Tsuboi}, M., {Miyazaki}, A., \& {Uehara}, K. 2015, \pasj, 67, 90

\bibitem[{{Utomo} {et~al.}(2015){Utomo}, {Blitz}, {Davis}, {Rosolowsky},
  {Bureau}, {Cappellari}, \& {Sarzi}}]{Utomo2015}
{Utomo}, D., {Blitz}, L., {Davis}, T., {et~al.} 2015, \apj, 803, 16

\bibitem[{{Valdivia} {et~al.}(2016){Valdivia}, {Hennebelle}, {G{\'e}rin}, \&
  {Lesaffre}}]{Valdivia2016}
{Valdivia}, V., {Hennebelle}, P., {G{\'e}rin}, M., \& {Lesaffre}, P. 2016,
  \aap, 587, A76

\bibitem[{{V{\'a}zquez-Semadeni} {et~al.}(2006){V{\'a}zquez-Semadeni}, {Ryu},
  {Passot}, {Gonz{\'a}lez}, \& {Gazol}}]{vazquezsemadeni2006}
{V{\'a}zquez-Semadeni}, E., {Ryu}, D., {Passot}, T., {Gonz{\'a}lez}, R.~F., \&
  {Gazol}, A. 2006, \apj, 643, 245

\bibitem[{{Walder} \& {Folini}(1998{\natexlab{a}})}]{Walder1998a}
{Walder}, R., \& {Folini}, D. 1998{\natexlab{a}}, \aap, 330, L21

\bibitem[{{Walder} \& {Folini}(1998{\natexlab{b}})}]{Walder1998b}
---. 1998{\natexlab{b}}, \apss, 260, 215

\bibitem[{{Williams} {et~al.}(2000){Williams}, {Blitz}, \&
  {McKee}}]{Williams2000}
{Williams}, J.~P., {Blitz}, L., \& {McKee}, C.~F. 2000, Protostars and Planets
  IV, 97

\bibitem[{{Williams} \& {McKee}(1997)}]{Williams1997}
{Williams}, J.~P., \& {McKee}, C.~F. 1997, \apj, 476, 166

\bibitem[{{Wolfire} {et~al.}(1995){Wolfire}, {Hollenbach}, {McKee}, {Tielens},
  \& {Bakes}}]{Wolfire1995}
{Wolfire}, M.~G., {Hollenbach}, D., {McKee}, C.~F., {Tielens}, A.~G.~G.~M., \&
  {Bakes}, E.~L.~O. 1995, \apj, 443, 152

\bibitem[{{Wolfire} {et~al.}(2003){Wolfire}, {McKee}, {Hollenbach}, \&
  {Tielens}}]{Wolfire2003}
{Wolfire}, M.~G., {McKee}, C.~F., {Hollenbach}, D., \& {Tielens}, A.~G.~G.~M.
  2003, \apj, 587, 278

\bibitem[{{Xu} {et~al.}(2016){Xu}, {Li}, {Yue}, \& {Goldsmith}}]{Xu2016}
{Xu}, D., {Li}, D., {Yue}, N., \& {Goldsmith}, P.~F. 2016, \apj, 819, 22

\bibitem[{{Yonekura} {et~al.}(1997){Yonekura}, {Dobashi}, {Mizuno}, {Ogawa}, \&
  {Fukui}}]{Yonekura1997}
{Yonekura}, Y., {Dobashi}, K., {Mizuno}, A., {Ogawa}, H., \& {Fukui}, Y. 1997,
  \apjs, 110, 21

\end{thebibliography}
